\documentclass[aps,prb,noshowpacs,twocolumn]{revtex4-2}
\usepackage{amsmath,amssymb,bm,dcolumn,epsfig,graphicx,longtable,xcolor,setspace,epsfig}
\usepackage[export]{adjustbox}
\begin{document}
\title{Density analysis for estimating the degree of on-site correlation on transition-metal atoms in extended systems}
\author{Leila Kalantari}
\author{Fabien Tran}
\author{Peter Blaha}
\affiliation{Institute of Materials Chemistry, Vienna University of Technology, Getreidemarkt 9/165-TC, A-1060 Vienna, Austria}
\begin{abstract}

In the context of the modified Becke-Johnson (mBJ) potential, we recently underlined that $\bar{g}$, the average of $\left\vert\nabla\rho\right\vert/\rho$ in the unit cell, has markedly different values in transition-metal oxides and pure transition metals [Tran \textit{et al}., J. Appl. Phys. \textbf{126}, 110902 (2019)]. However, since $\bar{g}$ is a constant it is not able to provide local information about a particular atom in the system. Furthermore, while $\overline{g}$ can be used only for periodic bulk solids, a local (i.e., position-dependent) version would allow us to consider also low-dimensional systems and interfaces. Such a local function has been proposed by Rauch \textit{et al}. [J. Chem. Theory Comput. \textbf{16}, 2654 (2020)] for the local mBJ potential. Actually, a local version of $\overline{g}$, or of another similar quantity like the reduced density gradient $\overline{s}$, could also be used in the framework of other methods. Here, we explored the idea to use such a local function $\tilde{g}$ (or $\tilde{s}$), defined as the average of $g$ (or $s$) over a certain region around a transition-metal atom, to estimate the degree of on-site correlation on this atom. We found a large difference in our correlation estimators between non-correlated and correlated materials, proving its usefulness and reliability. Our estimators can subsequently be used to determine whether or not a Hubbard $U$ on-site correction in the DFT+$U$ method should be applied to a particular atom. This is particularly interesting in cases where the degree of correlation of the transition-metal atoms is not clear, like interfaces between correlated and non-correlated materials or oxygen-covered metal surfaces. In such cases, our estimators could also be used for an interpolation of $U$ between correlated and non-correlated atoms.

\end{abstract}

\maketitle

\section{\label{introduction}Introduction}

Density functional theory (DFT) \cite{HohenbergPR64,KohnPR65} is the main computational tool to perform electronic structure calculations on systems of realistic complexity, however, one has to be careful when dealing with strongly correlated systems like Mott insulators. Prototypical such examples are the antiferromagnetic (AFM) $3d$-transition-metal oxides (TMO), for which DFT predictions using standard semilocal approximations for the exchange-correlation energy are very inaccurate (typically, a too small or even absent band gap, and a too small magnetic moment \cite{TerakuraPRB84}). More accurate DFT approximations, like hybrid functionals \cite{BredowPRB00,PerryPRB01,MuscatCPL01,HeydJCP03}, or beyond-DFT methods, like DFT plus dynamical mean field theory \cite{MetznerPRL89,KotliarRMP06,HeldAP07} or reduced density matrix functional theory \cite{SharmaPRL13,DiSabatinoJCP15}, can improve significantly the description of correlated systems, but they are prohibitively expensive for large systems. However, at a practical level there exist DFT methods that are barely more expensive than standard semilocal functionals, and therefore suitable for calculations on large systems, but also more reliable for systems with strongly correlated electrons. This is the case for instance of the modified Becke-Johnson (mBJ) potential \cite{TranPRL09,TranJPCA17} and DFT+$U$ \cite{AnisimovPRB93,YlvisakerPRB09,HimmetogluIJQC14}, both being as accurate as the hybrid functionals for strongly correlated systems.

In DFT+$U$, the Hubbard-model parameters to represent the on-site screened Coulomb (Hubbard $U$) and exchange (Hund $J$) interactions are used. $U$, which is usually larger than $J$, is a measure of the on-site interaction between two electrons in the same electronic shell and is defined as 
\begin{equation}
    U=E(d^{n+1})+E(d^{n-1})-2E(d^{n}).
    \label{eq:u}
\end{equation}
It is the Coulomb energy cost to transfer one electron from one site to the other such that the number of on-site interactions between two electrons is increased by one (from $n^{2}-n$ to $n^{2}-n+1$). In Eq.~(\ref{eq:u}), $E(d^{n})$ is the energy of an atom with $n$ localized electrons in the $d$ (or $f$) shell. In order to achieve reasonable results with DFT+$U$, two important points should be considered.

First, although a qualitative improvement of the results for correlated systems can be obtained by DFT+$U$, the results depend of course on the value of the parameters $U$ and $J$. Their values are often empirically calibrated such that the result for a property (e.g., band gap or oxidation energy) matches experiment. However, they can also be obtained \textit{ab initio} by some methods like constrained-LDA \cite{DederichsPRL84,HybertsenPRB89,MadsenEPL05}, constrained-RPA \cite{VaugierPRB2012,SpringerPRB98,AryasetiawanPRB2004,AryasetiawanPRB2006,SasiogluPRB2012}, or from linear response \cite{CococcioniPRB2005,PickettPRB98}. Nevertheless, there are still some ambiguity as well as freedom in the numerical implementation of these methods. For instance, localized $d$ or $f$ orbitals in solids usually hybridize with other valence $sp$ orbitals, causing an entangled band structure with considerable band widths, and it is difficult to uniquely define the localized states in solids.

Second, it is not even always clear, or known in advance, if $U$ should be applied or not. In contrast to AFM TMO where a large $U$ value of 6$-$8~eV is required, in a pure transition metal (TM) the $d$ electrons are itinerant, i.e. only weakly correlated, and no $U$ needs to be used in principle. However, there are of course intermediate cases, and furthermore in a given (complicated) system the degree of correlation may vary from one atom to the other even of the same type. This is the case, for instance, when a surface system consists of a TMO layer adsorbed on top of a pure metal; the degree of correlation on a TM atom is expected to decrease when going from the surface (TMO-like) deep into the bulk (pure TM-like).

In the present work we will show that quantities depending on the electron density $\rho$ (that we will call correlation estimators) can be used to distinguish between correlated and non-correlated $3d$ TM atoms in any kind of systems. Bulk solids, interfaces, and surfaces will be considered. Such a quantity could in principle be used to determine, at least qualitatively, whether or not a Hubbard $U$ correction should be applied on a TM atom.

We mention that Wang and Jiang proposed in Ref.~\cite{WangJCP19} to use $\rho$ to parameterize $U$ and $J$. More specifically, a simple average of $\rho$, calculated either in the whole unit cell or only in a sphere surrounding the atom, was used as screening parameter in the Slater integrals that enter into the expression of $U$ and $J$. They applied the method for calculating the energy difference between the AFM and ferromagnetic (FM) states of strongly correlated bulk solids.

The paper is organized as follows. In Sec.~\ref{theory}, the theory and the computational details are given. In Sec.~\ref{sec:application}, the results are presented and discussed, and in Sec.~\ref{summary} a summary is given.

\section{\label{theory}Theory and computational details}

\subsection{\label{correlationestimators}Correlation estimators}

\subsubsection{$\tilde{g}(\mathbf{r})$}

Our first correlation estimator is based on
\begin{equation}
g(\mathbf{r})=\frac{|\nabla\rho(\mathbf{r})|}{\rho(\mathbf{r})},
\label{eq:g}
\end{equation}
where $\rho$ is the electron density and $\nabla\rho$ the first derivative.
In Ref. \cite{KrukauJCP06}, it was proposed to use $g$ to calculate the screening parameter $\omega$ ($\omega\propto g$), which determines the separation of short- and longe-range exchange in the screened hybrid functional HSE \cite{HeydJCP03}. Then, the average of $g$ in the whole unit cell of volume $V_{\text{cell}}$,
\begin{equation}
 \bar{g}=\frac{1}{V_{\text{cell}}}\int\limits_{\text{cell}}g(\mathbf{r}')d^{3}r',
 \label{eq:gbar}
  \end{equation}
was used to define the parameter $c$ that specifies the relative weights of the two terms in the mBJ potential \cite{TranPRL09}. $c$ was parameterized as $c=\alpha+\beta\sqrt{\bar{g}}$, where $\alpha$ and $\beta$ are constants.
In Ref.~\cite{TranJAP19}, we discussed about the ability of $\bar{g}$ to distinguish between strongly correlated TMO and itinerant elemental TM by having clearly different values. However, since $\bar{g}$ is a constant for a particular system it can not be used as a local probe and distinguish between different atoms in the same system. Furthermore, on the technical side it is not applicable to non-periodic solids, interfaces, and systems with vacuum (low-dimensional systems), since in such systems averaging a quantity in the unit cell has no meaning.

In the present work, we are searching for a quantity that is local (i.e., position dependent) and can provide an indication about the strength of correlation on a particular atom. Using simply Eq.~(\ref{eq:g}) would not really work since some (local) average (as in $\bar{g}$) is still necessary in order to have a function that is able as $\bar{g}$ to distinguish between TMO and pure TM (see discussion in Sec.~\ref{bulk}). To this end we will use the smeared local estimator first suggested in Ref.~\cite{MarquesPRB11} and then implemented by Rauch \textit{et al}. \cite{RauchJCTC2020,RauchPRB20}. This local correlation estimator, $\tilde{g}$, is a local average of $g$:
\begin{equation}
\tilde{g}(\mathbf{r})=\frac{1}{(2\pi\sigma^2)^{3/2}}\int  g(\mathbf{r}')e^{-\frac{|\mathbf{r}-\mathbf{r}'|^2}{2\sigma^2}}d^{3}r',
\label{eq:gtilde}
\end{equation}
where the smearing parameter $\sigma$ determines the size of the region (centered around $\mathbf{r}$) over which $g$ is averaged. The details of how $\sigma$ can be determined and set as a parameter will be discussed in Sec.~\ref{sec:application}. The expression for $\tilde{g}$ is very advantageous since it can be very easily calculated by using the convolution theorem if $g$ is expanded in plane waves. Note that in periodic bulk systems, $\tilde{g}$ becomes a constant and recovers the value of Eq.~(\ref{eq:gbar}) when $\sigma$ is large enough.

However, for surfaces and other systems with vacuum numerical issues with $|\nabla\rho|/\rho$ that becomes very large close to the surface region need to be solved. We followed the prescription proposed in Ref.~\cite{RauchJCTC2020}, which consists of modifying Eq.~(\ref{eq:g}) as follows:
\begin{equation}
    g(\mathbf{r})=\frac{1-\alpha}{\beta}\bigg[1-\text{erf}\bigg(\frac{\rho(\mathbf{r})}{\rho_{\text{th}}}\bigg)\bigg]+\frac{|\nabla\rho(\mathbf{r})|}{\rho(\mathbf{r})}\text{erf}\bigg(\frac{\rho(\mathbf{r})}{\rho_{\text{th}}}\bigg),
\label{eq:g2}
\end{equation}
where $\rho_{\text{th}}$ is a threshold for very low densities.
For $\rho \gg \rho_{\text{th}}$, $|\nabla\rho|/\rho$ is obtained, while for $\rho \ll \rho_{\text{th}}$, $g$ becomes $\left(1-\alpha\right)/\beta=\left(1-(-0.012)\right)/1.023=0.989$. 
Equation~(\ref{eq:g2}) was proposed in the framework of the local mBJ potential to cope with the aforementioned problem, but also to have the correct asymptotic behavior of the local mBJ potential in the vacuum region. Here, the goal of the damping with the function $\text{erf}(\rho/\rho_{\text{th}})$ is more to have a faster convergence of the plane-wave expansion of $g$. Although the first term in Eq.~(\ref{eq:g2}) is not necessary and could be discarded for the purpose of the present work, we decided to use the full original expression from Ref.~\cite{RauchJCTC2020}. How $\rho_{\text{th}}$ is chosen in the present work will be explained in Sec.~\ref{sec:application}.

\subsubsection{$\tilde{s}(\mathbf{r})$}

The second correlation estimator that we will consider is based on the reduced density gradient $s$, which is used in the enhancement factor of exchange functionals of the generalized gradient approximation (GGA) \cite{PerdewPRL96}. $s$ reads
\begin{equation}
    s(\mathbf{r})=\frac{|\nabla\rho(\mathbf{r})|}{2(3\pi^{2})^{1/3}\rho^{4/3}(\mathbf{r})}.
    \label{eq:s}
\end{equation}
Besides a constant factor, $s$ differs from Eq.~(\ref{eq:g}) by the power in the denominator that makes $s$ dimensionless. Far from the nuclei, $s$ goes to infinity (while $g$ goes to a constant), but does not show the large values close to surface regions like $g$. Similarly to the second term in Eq.~(\ref{eq:g2}) for $g$, we will damp $s$ in the vacuum:
\begin{equation}
    s(\mathbf{r})=\frac{|\nabla\rho(\mathbf{r})|}{2(3\pi^{2})^{1/3}\rho^{4/3}(\mathbf{r})}\text{erf}\bigg(\frac{\rho(\mathbf{r})}{\rho_{\text{th}}}\bigg)
    \label{eq:s2}
\end{equation}
and then use it in
\begin{equation}
\tilde{s}(\mathbf{r})=\frac{1}{(2\pi\sigma^2)^{3/2}}\int  s(\mathbf{r}')e^{-\frac{|\mathbf{r}-\mathbf{r}'|^2}{2\sigma^2}}d^{3}r'
\label{eq:stilde}
\end{equation}
to get our second correlation estimator.

\subsection{\label{methods}Computational details}

All calculations presented in this work were carried out using the all-electron WIEN2k code \cite{WIEN2k,BlahaJCP20}, which is based on the full potential linearized augmented plane-wave and local orbitals [FP-(L)APW+lo] method \cite{Singh,KarsaiCPC17}. In the FP-(L)APW+lo method, the wave functions, electron density, and potential are expanded in spherical harmonics inside the atomic spheres and in plane waves in the interstitial region. The GGA functional of Perdew, Burke, and Ernzerhof (PBE) \cite{PerdewPRL96} has been used to treat the exchange-correlation effects in the self-consistent calculations. The parameters of the calculations, like the number of $\mathbf{k}$-points or the size of the basis set, have been chosen such that the results are well converged. Typically, we used a basis-set size corresponding to $R_{\text{MT}}^{\text{min}}K_{\text{max}}=9$, where $R_{\text{MT}}^{\text{min}}$ is the smallest atomic sphere radius in the system and $K_{\text{max}}$ is the magnitude of the largest reciprocal lattice vector.

An important point concerning the calculation of the correlation estimators $\tilde{g}$ and $\tilde{s}$ is the following. WIEN2k is an all-electron code, therefore a plane-wave expansion in the entire unit cell of the wave functions and all derived quantities like $\rho$ is in principle practically impossible. However, $g$ and $s$, and more particularly the damped versions Eqs.~(\ref{eq:g2}) and (\ref{eq:s2}), are smooth enough such that it is possible to expand them in plane waves in the entire unit cell with a Fourier series that is not prohibitively large. Therefore, the calculation of the correlation estimators $\tilde{g}$ and $\tilde{s}$ can be performed efficiently using the convolution theorem. If the convolution theorem could not be used, the calculation of $\tilde{g}$ and $\tilde{s}$ would be very cumbersome (see discussion in Ref.~\cite{SinghPRB93} for the weighted-density approximation).

\section{\label{sec:application}Applications}

\subsection{\label{bulk}Bulk solids}

\begin{figure}
    \includegraphics[width=\columnwidth]{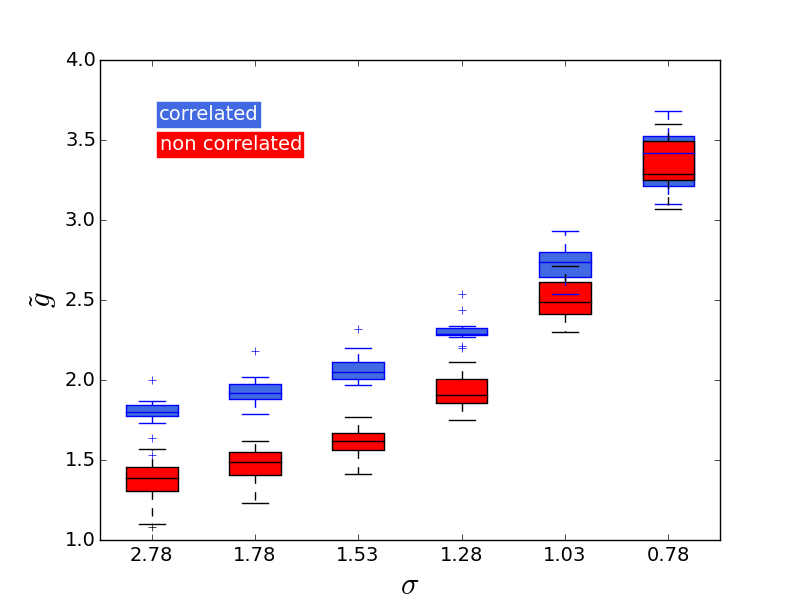}
    \includegraphics[width=\columnwidth]{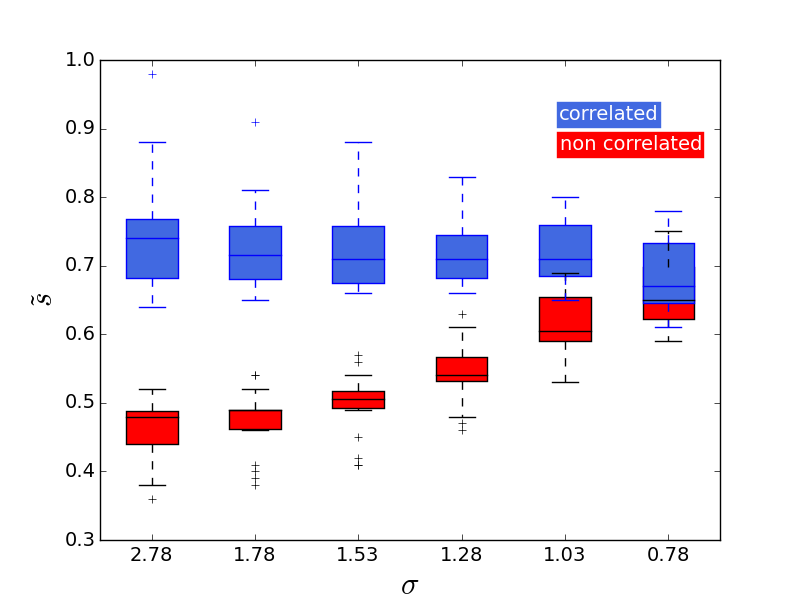}
    \caption{Box plot of the values of $\tilde{g}$ (upper panel) and $\tilde{s}$ (lower panel) on the $3d$ TM atom in correlated and non-correlated systems obtained with different values of the smearing parameter $\sigma$ (in bohr). The inner horizontal line in each box shows the median and the outer horizontal lines are the minimum and maximum values, while the outliers are shown with plus signs.}
     \label{All3d_gs}
\end{figure}

\begin{table*}
    \caption{\label{results}Values of $\tilde{g}$ and $\tilde{s}$ on the $3d$ TM atom in correlated and non-correlated bulk solids calculated with $\sigma=1.78$~bohr. The magnetic state is indicated in parenthesis.}
\begin{ruledtabular}
    \begin{tabular}{lcclcc}
Non-correlated & $\tilde{g}$ & $\tilde{s}$ & Correlated & $\tilde{g}$ & $\tilde{s}$ \\
\hline
Ti (NM)              &1.23&0.41 & TiO$_{2}$-anatase (NM)&2.02&0.81\\ 
V (NM)               &1.27&0.39 & TiO$_{2}$-rutile (NM) &1.98&0.76\\
Cr (AFM)              &1.35&0.40 & Ti$_{2}$O$_{3}$ (NM)   &1.90&0.71\\
Mn (NM)              &1.40&0.47 & V$_{2}$O$_{3}$ (AFM)    &1.93&0.70\\
Fe (FM)              &1.43&0.48 & SrVO$_{3}$ (NM)        &1.90&0.68\\	    
FeAl (FM)            &1.23&0.38 & Cr$_{2}$O$_{3}$ (AFM)   &1.97&0.75\\ 
FeNi (FM)            &1.50&0.49 & CrO$_{2}$ (FM)	        &2.00&0.72\\ 
Fe$_{3}$Ni (FM)      &1.48&0.49 & MnO (AFM)	            &1.85&0.66\\ 
Fe$_{2}$P (FM)       &1.42&0.46 & MnO$_{2}$ (AFM) &2.02 & 0.75\\
FeSb$_{2}$(FM)       &1.50&0.52& Mn$_{2}$O$_{3}$ (AFM)   &1.99&0.72\\		   
Co (FM)              &1.51&0.49 & FeO (AFM)	        &1.83&0.65 \\ 
Ni (FM)              &1.57&0.49 & Fe$_{2}$O$_{3}$ (AFM)	&1.96&0.73\\ 
Cu (NM)              &1.58&0.49 & Fe$_{3}$O$_{4}$ (FM)	&1.92&0.70\\ 
Cu$_{2}$Sb (NM)      &1.52&0.49 & FeF$_{2}$ (AFM)	        &2.18&0.91\\	     
Cu$_{3}$P (NM)       &1.56&0.54 & CoO (AFM)	        &1.86&0.66\\ 
CuAu (NM)            &1.61&0.52 & NiO (AFM)	        &1.90&0.66\\ 
Cu$_{3}$Au (NM)      &1.62&0.54 & CuO (AFM)	        &1.91&0.77\\    
Zn (NM)              &1.47&0.49 & Cu$_{2}$O (NM) 	&1.79&0.68\\
&&& CuI (NM) &1.79&0.79\\
&&&ZnO (NM)	         &1.97&0.80\\  
&&&YBa$_{2}$Cu$_{3}$O$_{6}$ (FM) - Cu1,Cu2&1.81,1.89&0.78,0.70\\ 
&&&YBa$_{2}$Cu$_{3}$O$_{7}$ (NM) - Cu1,Cu2&1.81,1.89&0.77,0.70\\
\end{tabular}   
\end{ruledtabular}
\end{table*}

In order to find a relation between the degree of on-site correlation and $\tilde{g}$ or $\tilde{s}$ on a $3d$ TM atom, we calculated these two correlation estimators for a series of bulk solids with a wide range of correlation strength: from pure TM (itinerant non- or weakly correlated) to AFM oxides (strongly correlated). All considered systems are listed in Table~\ref{results} along with their magnetic state: non-magnetic (NM), FM, or AFM. For the non-/weakly correlated systems, not only pure TM are considered, but also other systems such as FeAl or Cu$_{3}$P. For the more correlated systems, AFM and NM oxides are considered, including YBa$_{2}$Cu$_{3}$O$_{6}$ and its parent compound YBa$_{2}$Cu$_{3}$O$_{7}$, which is a high-$T_{c}$ superconductor.

Figure~\ref{All3d_gs} shows the calculated $\tilde{g}$ and $\tilde{s}$ on the TM atoms for different smearing parameters $\sigma$ in Eqs.~(\ref{eq:gtilde}) and (\ref{eq:stilde}), respectively. The largest considered value of $\sigma$ is 3.78~bohr (2~\AA), as originally chosen in Ref.~\cite{RauchJCTC2020} for the local mBJ potential, so that $\tilde{g}$ and $\tilde{s}$ are averaged over a region that covers typical interatomic distances. Then we gradually reduced $\sigma$ until 0.78~bohr, which is more representative of the size of an atom. Since the results for $\sigma=3.78$ and 2.78~bohr lead to the same results for the bulk systems, we do not show the values for $\sigma=3.78$~bohr. The main observations that can be made are the following. The most interesting one is that except for the smallest values of $\sigma$ (0.78 and 1.03~bohr), the values of the correlation estimators in correlated and non-correlated systems are clearly different. For a value of $\sigma$ that is too small, the average is done only around the atomic core region and the resulting value becomes system-independent and no distinction between correlated and non-correlated systems can be made. We can also observe that the range of values of $\tilde{g}$ is for the correlated systems quite narrow. Also, $\tilde{g}$ shows a pronounced variation with respect to $\sigma$ and increases when $\sigma$ decreases. On the other hand, for the correlated systems $\tilde{s}$ has a much larger spread than $\tilde{g}$ and is quite independent of $\sigma$. In any case, both $\tilde{g}$ and $\tilde{s}$ seem to be good candidates to be used as correlation estimators. In the rest of this section, the discussion will be mainly based on the results obtained with $\sigma=1.78$~bohr.

The numerical values of $\tilde{g}$ and $\tilde{s}$ obtained with $\sigma=1.78$~bohr are shown in Table~\ref{results}. For the non-correlated systems the values of $\tilde{g}$ range from 1.2 to 1.6, while they are clearly larger for the correlated systems, between 1.8 and 2.2. For $\tilde{s}$, the range is 0.4$-$0.55 for the non-correlated systems and 0.65$-$0.9 for the correlated solids. Thus, for instance $\tilde{g}=1.7$ and $\tilde{s}=0.6$ for $\sigma=1.78$ could be used as boundary values to distinguish between correlated and non-correlated systems, and to decide if, at least qualitatively, a particular TM atom would need or not a Hubbard on-site correction $U$ in a DFT+$U$ calculation. However, it seems to be difficult to estimate a value of $U$ from the estimators, because we do not really see any systematic trend neither across the $3d$ series nor within a certain class of compounds.

Among the correlated solids, we note that FeF$_{2}$ and CuI are not oxides. While the former is a typical correlated AFM solid, the latter, which has the zinc blend structure, is a highly mobile $p$-type wide band gap nonmagnetic semiconductor (similar as ZnO or TiO$_2$). Considering it as correlated is somehow consistent with the results from Refs.~\cite{JiangJCP13,RUBELCPC21}, where it is shown that the mBJ potential alone is not accurate enough to yield the experimental band gap, so that adding an effective Hubbard term $U$ is necessary.


\begin{table*}
\caption{Values of $\tilde{g}$ and $\tilde{s}$ on the Ni atom in FM fcc Ni and AFM NiO calculated with different smearing parameters $\sigma$ (in bohr).}
\begin{ruledtabular}
\begin{tabular}{lcccccc}
$\sigma$ & 2.78 & 1.78 & 1.53 & 1.28 & 1.03 & 0.78\\
\hline
\multicolumn{1}{l}{} & \multicolumn{6}{c}{$\tilde{g}$} \\ 
Ni      & 1.52 & 1.57 & 1.67 & 1.95 & 2.50 & 3.37\\
NiO     & 1.81 & 1.90 & 2.03 & 2.29 & 2.77 & 3.51\\
\hline
\multicolumn{1}{l}{} & \multicolumn{6}{c}{$\tilde{s}$} \\
Ni     & 0.48 & 0.49 & 0.50 & 0.54 & 0.60 & 0.65 \\ 
NiO    & 0.65 & 0.66 & 0.67 & 0.70 & 0.72 & 0.71 \\
\end{tabular}
\end{ruledtabular}
\label{Ta:Ni-NiO}
\end{table*}

\begin{figure}
    \centering
    \includegraphics[width=\columnwidth]{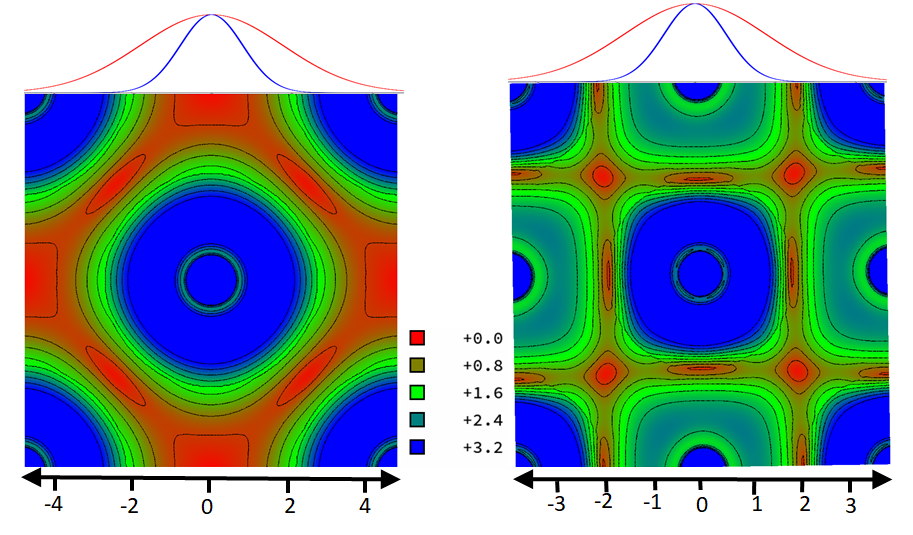}
    \caption{Two-dimensional plots of $g$ in FM fcc Ni (left) and AFM NiO (right) in a (001) plane. A Ni atom is always at the center. The scale of the ruler is in bohr and on top Gaussian functions for $\sigma=0.78$ (blue) and 1.78 (red) bohr are shown.}
    \label{Ni_NiO_GRR}
\end{figure}
\begin{figure}
    \centering
    \includegraphics[width=\columnwidth]{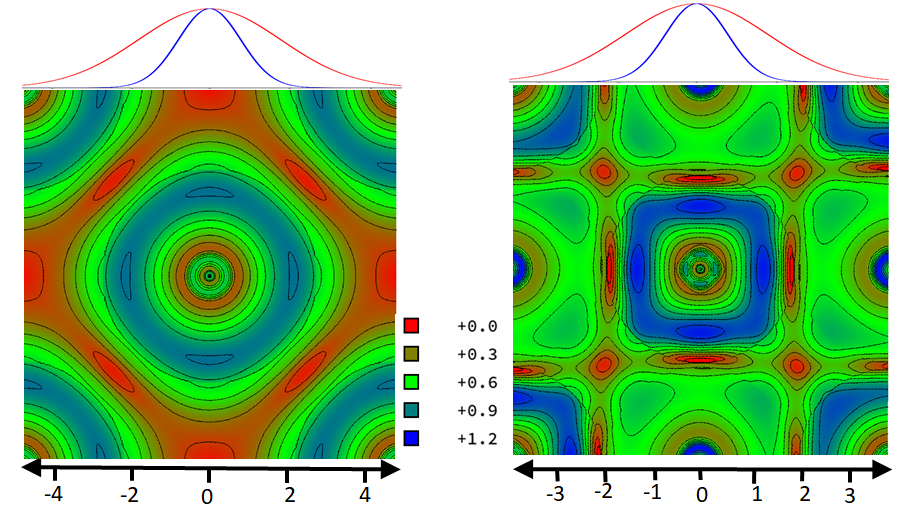}
    \caption{Two-dimensional plots of $s$ in FM fcc Ni (left) and AFM NiO (right) in a (001) plane. A Ni atom is always at the center. The scale of the ruler is in bohr and on top Gaussian functions for $\sigma=0.78$ (blue) and 1.78 (red) bohr are shown.}
    \label{Ni_NiO_S}
\end{figure}

In order to discuss about $\tilde{g}$ and $\tilde{s}$ in more detail, and to show the effect of $\sigma$, we take a closer look at two examples: FM face-centered cubic (fcc) Ni and AFM NiO (in rocksalt structure) as non-correlated and correlated systems, respectively. Figures~\ref{Ni_NiO_GRR} and \ref{Ni_NiO_S} show two-dimensional plots of $g$ and $s$ for Ni and NiO within a (001) plane with a Ni atom at the center of the plot. Gaussian functions with $\sigma=0.78$ and 1.78~bohr are added to the figures to show which area is covered by the integration in Eqs.~(\ref{eq:gtilde}) and (\ref{eq:stilde}). With $\sigma=0.78$~bohr the integration is done only in the region close to the Ni atom, thus giving values of $\tilde{g}$ and $\tilde{s}$ in Ni and NiO that are rather close (see Table~\ref{Ta:Ni-NiO}). With $\sigma=1.78$~bohr the low $s$ and $g$ area is also covered, which leads to more different values for $\tilde{g}$ and $\tilde{s}$ in non-correlated and correlated systems, as discussed above. Table~\ref{Ta:Ni-NiO} shows the values of $\tilde{g}$ and $\tilde{s}$ in Ni and NiO for other values of $\sigma$. In order to sufficiently take into account the environment of the Ni atom (i.e., for NiO the effect due to the oxygen atoms) for calculating the average, $\sigma$ has to be large enough, let us say at least 1.2~bohr.

\subsection{Interfaces and surfaces}

Moving to more complex systems, the values of $\tilde{g}$ and $\tilde{s}$ on TM atoms in interface and surface systems are discussed below.

\subsubsection{Interfaces}

\begin{figure}
    \centering
     \includegraphics[width=\columnwidth]{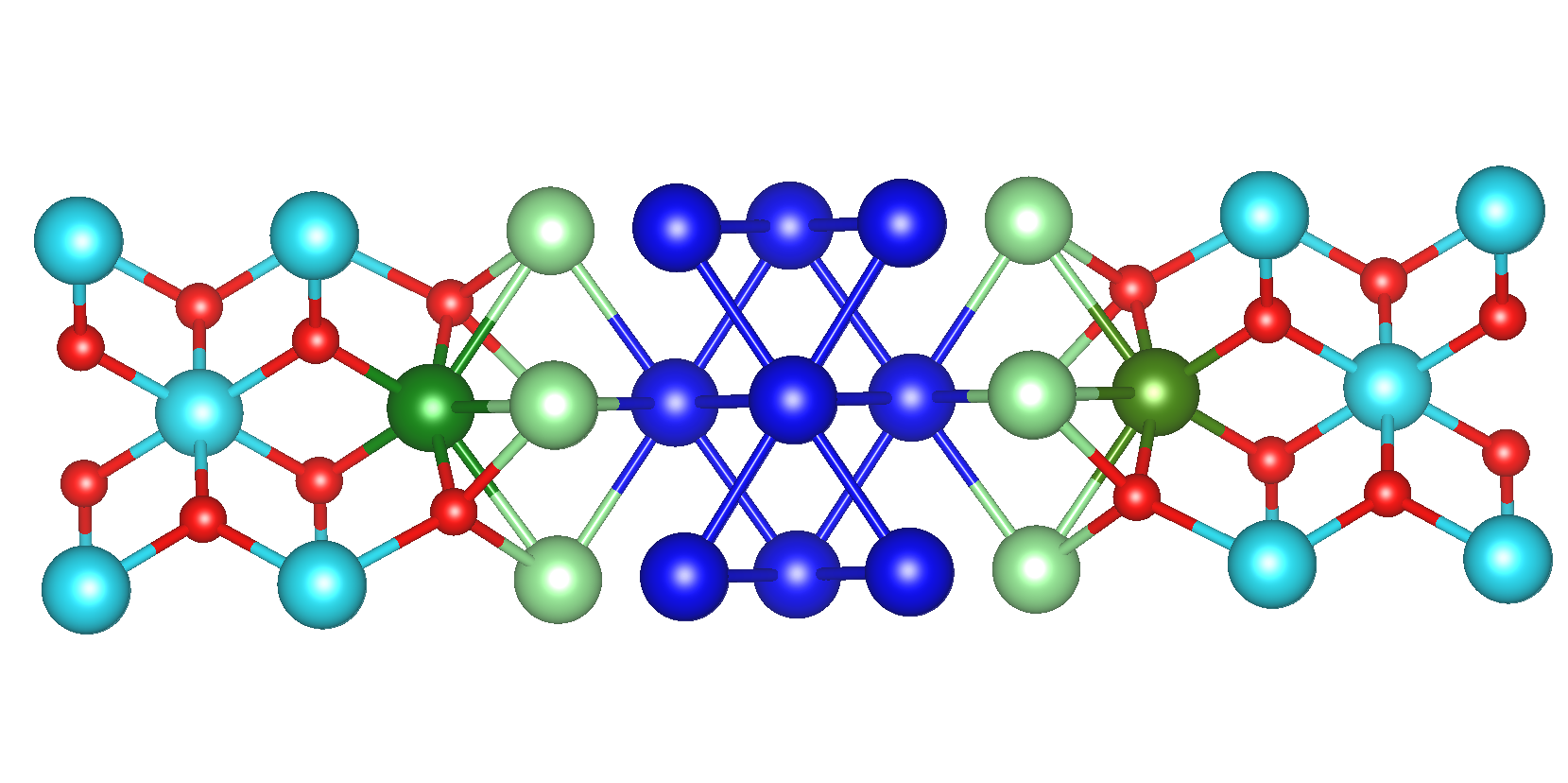}
     \caption{Side view of the Mn/MnO$_{2}$ interface system. The color coding of the atoms is as follows: red (O), dark blue (Mn with no O neighbors), light green (Mn with two O neighbors), dark green (Mn with four O neighbors), light blue (Mn with six O neighbors).}
    \label{Mn-MnO2}
\end{figure}

\begin{figure}
    \centering
     \includegraphics[width=\columnwidth]{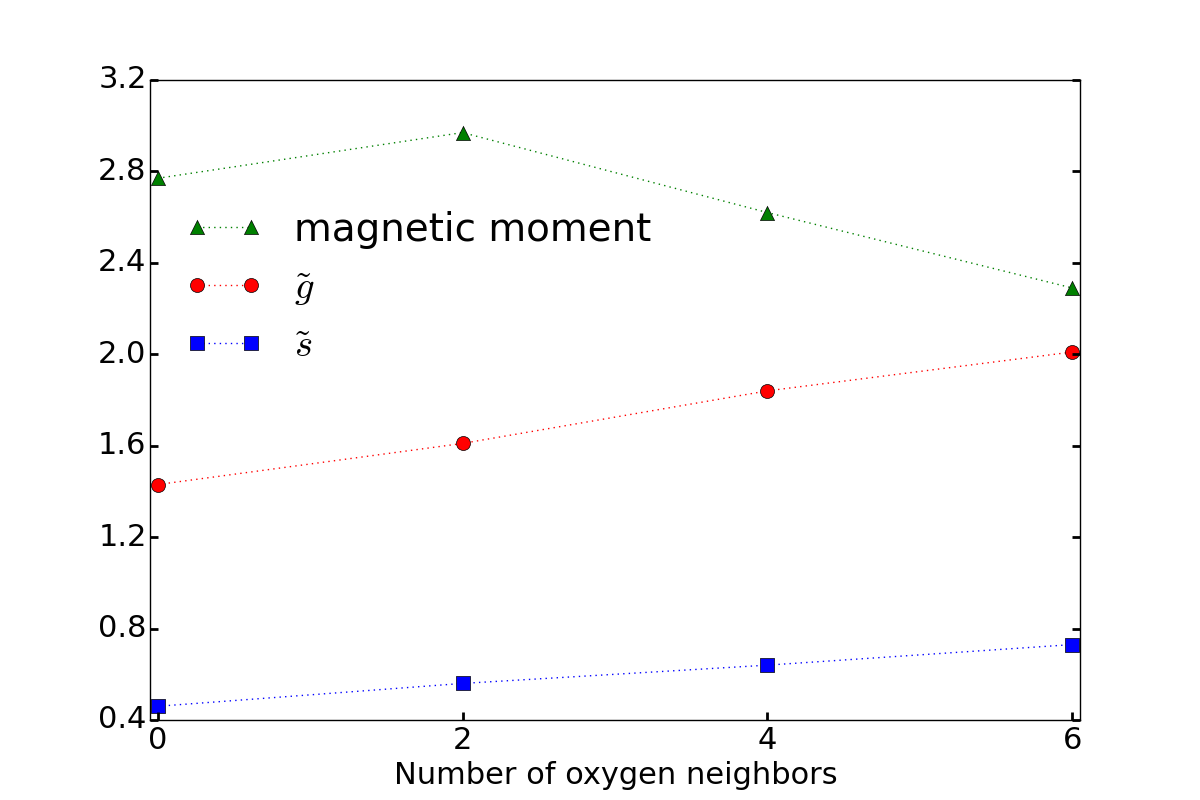}
     \caption{$\tilde{g}$, $\tilde{s}$, and magnetic moment $M$ ($\mu_B$) on Mn atoms in the Mn/MnO$_{2}$ interface system calculated with $\sigma=1.78$~bohr plotted against the number of oxygen atoms surrounding a Mn atom.}
    \label{Mn-MnO2-s-grr}
\end{figure}

We calculated $\tilde{g}$ and $\tilde{s}$ on Mn atoms in an interface of AFM body-centered cubic (bcc) Mn with MnO$_{2}$ in rutile structure. The (001) interface model consists of 5 Mn and 7 MnO$_2$ layers and has only a small $3.5\%$ lattice mismatch. As shown in Fig.~\ref{Mn-MnO2}, the interface model consists of Mn atoms in four different situations; one in the middle of the Mn slab without oxygen neighbors, one in the middle of the MnO$_{2}$ slab with six oxygen neighbors, and two in the interface region with two and four oxygen neighbors, respectively. Table~\ref{Ta:Mn-MnO} shows $\tilde{g}$ and $\tilde{s}$ on the Mn atoms obtained with different values of the smearing parameter $\sigma$. There is a clear relation between the correlation estimators and the oxygen coordination, except for small values of $\sigma$ (0.78~bohr for $\tilde{g}$ and up to 1.28~bohr for $\tilde{s}$). For comparison, $\tilde{g}$ and $\tilde{s}$ obtained from bulk calculations of Mn and MnO$_{2}$ (both AFM) are also shown in Table~\ref{Ta:Mn-MnO}. In most cases, the bulk values agree very well with the values on the corresponding side of the interface system [Mn(no O neighbor) and Mn(6 O neighbors), respectively]. However, some differences can also be noted like in the case of $\tilde{g}$ with $\sigma=3.78$~bohr for Mn or $\tilde{s}$ with the small values of $\sigma$ for MnO$_{2}$. Figure~\ref{Mn-MnO2-s-grr} shows the perfectly linear relation between our correlation estimators and the oxygen coordination number for $\sigma=1.78$~bohr. Thus with our correlation thresholds of 1.7 for $\tilde{g}$ or 0.6 for $\tilde{s}$ we can classify the Mn atoms with two oxygen neighbors still as non-correlated, while those with four oxygen neighbors should already be considered as correlated. As an alternative interpretation, one could also use $\tilde{g}$ or $\tilde{s}$ to interpolate the value of $U$ between 0 and 5~eV (assuming a $U$ of 5~eV for MnO$_2$). In Figure~\ref{Mn-MnO2-s-grr} we also show the magnetic moment of the Mn atoms. Metallic Mn has a larger moment than MnO$_2$, but the non-monotonic behavior indicates that the magnetic moment cannot be used as a correlation estimator. 

\begin{table*}
\caption{Values of $\tilde{g}$ and $\tilde{s}$ on Mn atoms in the Mn/MnO$_{2}$ interface system calculated with different values of $\sigma$ (in bohr). The values in bold for $\sigma$=1.78 indicate the correlated atoms. $\tilde{g}$ and $\tilde{s}$ in bulk Mn and MnO$_{2}$ (both AFM) are also shown for comparison.}
\begin{ruledtabular}
\begin{tabular}{lcccccc}
$\sigma$ & 2.78 & 1.78 & 1.53 & 1.28 & 1.03 & 0.78\\
\hline
\multicolumn{1}{l}{} & \multicolumn{6}{c}{$\tilde{g}$} \\ 
Mn(bulk)      & 1.30 & 1.40 & 1.57 & 1.91 & 2.49 & 3.28\\
Mn(no O neighbor) &1.36&1.43&1.58&1.89&2.46&3.25  \\ 
Mn(2 O neighbors) &1.53&1.61&1.76&2.05&2.56&3.29  \\
Mn(4 O neighbors) &1.72&\textbf{1.84}&1.96&2.19&2.60&3.25   \\
Mn(6 O neighbors) &1.88&\textbf{2.01}&2.12&2.32&2.67&3.27   \\
MnO$_{2}$(bulk)     & 1.89 & \textbf{2.02} & 2.13 & 2.33 & 2.69 & 3.29\\
\hline
\multicolumn{1}{l}{} & \multicolumn{6}{c}{$\tilde{s}$} \\
Mn(bulk)      & 0.44 & 0.47 & 0.50 & 0.57 & 0.68 & 0.75\\
Mn(no O neighbor) &0.45&0.46&0.50&0.56&0.66&0.73  \\ 
Mn(2 O neighbors) &0.55&0.56&0.59&0.64&0.72&0.76  \\
Mn(4 O neighbors) &0.66&\textbf{0.64}&0.63&0.63&0.63&0.60  \\
Mn(6 O neighbors) &0.75&\textbf{0.73}&0.71&0.69&0.66&0.62  \\
MnO$_{2}$(bulk)  & 0.77 & \textbf{0.75} & 0.75 & 0.74 & 0.74 & 0.72\\
\end{tabular}
\end{ruledtabular}
\label{Ta:Mn-MnO}
\end{table*}

\subsubsection{Surfaces}

\begin{figure}
    \centering
    \includegraphics[width=\columnwidth]{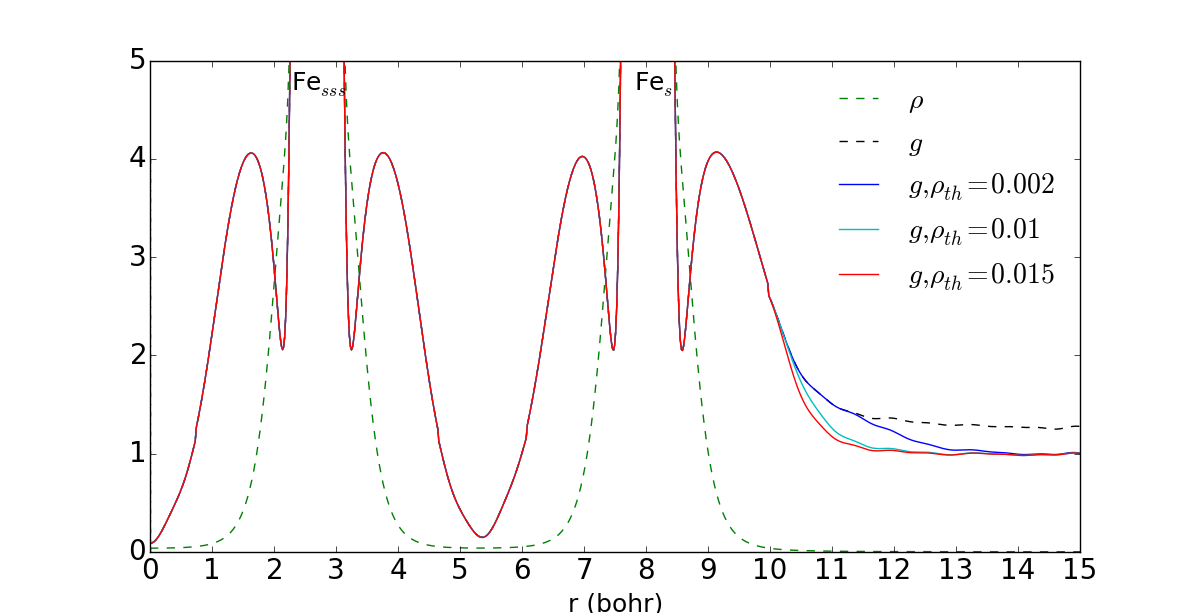}
    \caption{Electron density $\rho$, original $g$ [Eq.~(\ref{eq:g})], and $g$ damped with different threshold densities $\rho_{\text{th}}$ [Eq.~(\ref{eq:g2})] plotted for the Fe(001) surface along the [001] direction into the vacuum. Fe$_{sss}$ and Fe$_{s}$ indicate the  sub-subsurface and surface Fe atoms, respectively.}
    \label{Fe001O_grr}
\end{figure}

\begin{figure}
    \centering
    \includegraphics[width=\columnwidth]{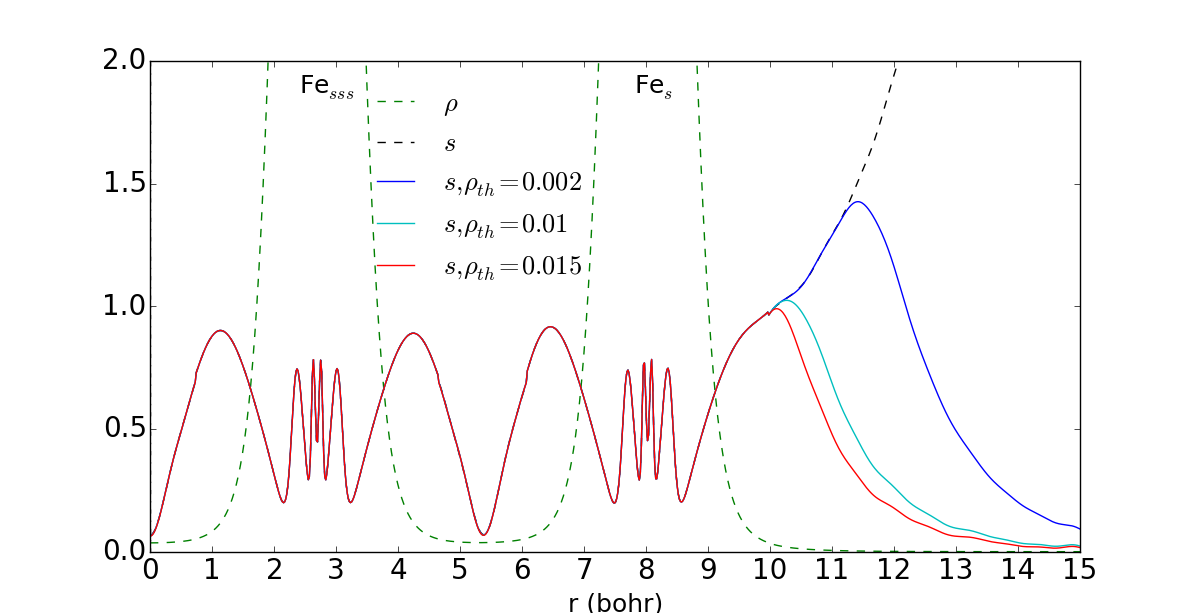}
    \caption{Electron density $\rho$, original $s$ [Eq.~(\ref{eq:s})], and $s$ damped with different threshold densities $\rho_{\text{th}}$ [Eq.~(\ref{eq:s2})] plotted for the Fe(001) surface along the [001] direction into the vacuum. Fe$_{sss}$ and Fe$_{s}$ indicate the  sub-subsurface and surface Fe atoms, respectively.}
    \label{Fe001O_SS}
\end{figure}

As mentioned in Sec.~\ref{correlationestimators} and in Ref.~\cite{RauchJCTC2020}, $g$ as given by Eq.~(\ref{eq:g}) is quite large in the vacuum region, while $s$ [Eq.~(\ref{eq:s})] even goes to infinity far from the nuclei. Therefore, they were damped by multiplying them by $\text{erf}(\rho/\rho_{\text{th}})$ \cite{RauchJCTC2020}, see Eqs.~(\ref{eq:g2}) and (\ref{eq:s2}). The threshold density $\rho_{\text{th}}$ is a parameter that has to be chosen suitably, i.e., not too small so that one has damping of $g$ and $s$ in the vacuum, but also not too large so that they are not affected inside the bulk. In fact for our purpose, $\rho_{\text{th}}$ should be chosen such that the values of $\tilde{g}$ and $\tilde{s}$ on the surface atoms are the same/similar as the values on atoms deeper in the bulk. Taking the Fe-(001) surface as example, the effect of the damping on $g$ and $s$ is displayed in Figs.~\ref{Fe001O_grr} and \ref{Fe001O_SS}, respectively. The figures show the electron density $\rho$ of sub-subsurface and surface Fe atoms and the corresponding $g$ or $s$, undamped and damped with $\rho_{\text{th}}$ = 0.002, 0.01, or 0.015~e/bohr$^3$.


We calculated $\tilde{g}$ and $\tilde{s}$ on the Ni and Fe atoms in surface systems with and without oxygen coverage. Different values of $\rho_{\text{th}}$ and $\sigma$ were considered, but
we show in Table~\ref{Ni-Fe-surface:Ta} only the results for $\tilde{g}$ and $\tilde{s}$ obtained with $\rho_{\text{th}}=0.015$~e/bohr$^{3}$ and $\sigma=2.78$~bohr. With this choice the correlation estimators for the plain surfaces are only very little enhanced as compared to the bulk, and the method is optimally sensitive to distinguish correlated and non-correlated atoms.

We first discuss Ni(111) surfaces. In the case of (full) oxygen coverage, the oxygen atom is located at the fcc hollow site. From Table~\ref{Ni-Fe-surface:Ta} we can see that for the plain Ni(111) surface the correlation estimators are very close to the bulk values. However, with full oxygen coverage the Ni atoms at the surface have values of $\tilde{g}$ and $\tilde{s}$ that are clearly larger than for the Ni atoms in the subsurface and deeper into the bulk. 
By comparing $\tilde{g}$ and $\tilde{s}$ with the corresponding values of bulk Ni and NiO we clearly can classify the surface Ni atom in the fully O covered surface as correlated.

\begin{figure}
    \centering
     \includegraphics[width=\columnwidth]{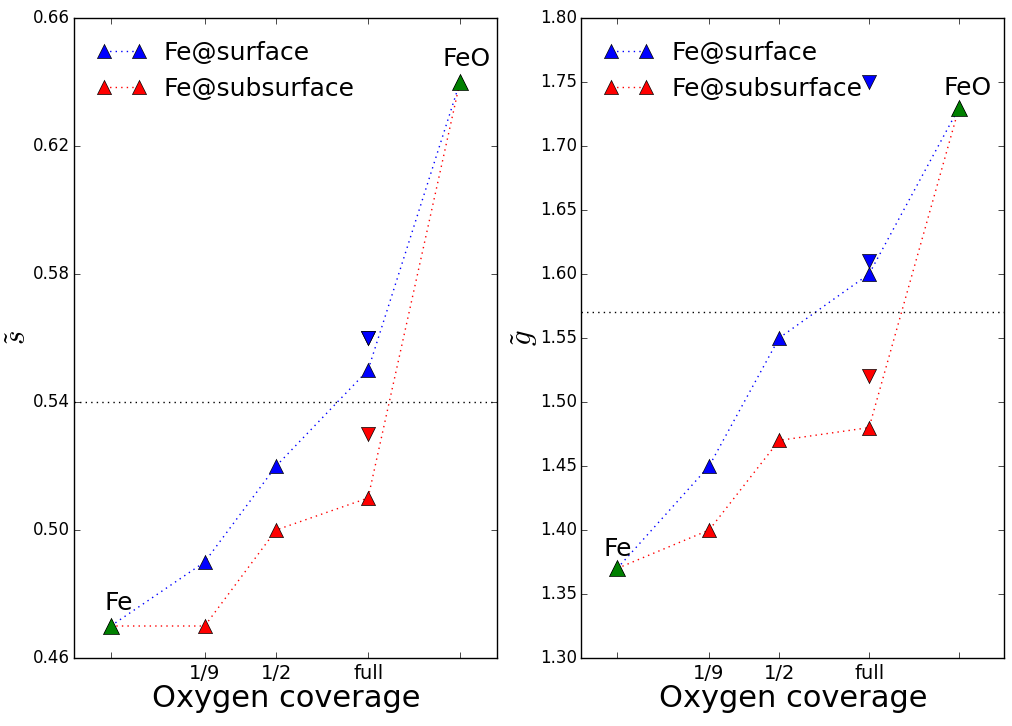}
     \caption{$\tilde{g}$ and $\tilde{s}$ for different Fe atoms on the surface and subsurface of Fe(001) surfaces with different O coverage calculated with $\sigma=2.78$~bohr. The down triangle are the results for the Fe(001) surface  with full oxygen coverage and one extra oxygen in the subsurface. The dashed line shows our selected value to separate correlated and non-correlated atoms. The green triangles are $\tilde{g}$ and $\tilde{s}$ for bulk Fe and bulk FeO.}
    \label{Fe_surface-S-grr}
\end{figure}

For Fe(001) surfaces, we studied plain Fe(001) and Fe(001) with different oxygen coverages, namely full, full with one additional oxygen atom in the subsurface (in a $2 \times 2$ supercell), half, and 1/9 ($3 \times 3$ supercell). In these cases the surface Fe atoms have 4, 4 or 5, 2 and 1 oxygen neighbors, respectively. The results in Table~\ref{Ni-Fe-surface:Ta} show that for both full oxygen coverages all atoms on the surface have the largest values of $\tilde{g}$ and $\tilde{s}$. In the case of half coverage $\tilde{g}$ and $\tilde{s}$ on the Fe atoms at the surface are smaller, whereas a further reduction is obtained for the Fe atoms in the subsurfaces (except in the case of full coverage(2)) and for all Fe atoms in the case of 1/9 coverage. In the subsurface of full coverage(2) each Fe atom has two oxygen nearest neighbors and is therefore expected to be more correlated. For comparison we also show the values for bulk Fe and FeO. For convenience we also show the results for the surface and subsurface Fe atoms graphically in Fig.~\ref{Fe_surface-S-grr} and identify the surface atoms in the full coverage case as correlated. Whether the Fe atoms on the surface with half oxygen coverage or in the subsurface with full oxygen coverage should be considered as correlated or not depends on the chosen boundary value (e.g., see dashed line in Fig.~\ref{Fe_surface-S-grr} as a possible choice) or one could use our estimators to interpolate the value of $U$ for the different Fe atoms. 

\begin{table}
    \caption{$\tilde{g}$ and $\tilde{s}$ on TM atoms in plain Ni(111) and Fe(001) surfaces, as well as in Ni(111) with full oxygen coverage and Fe(001) with different oxygen coverage. The values in bold indicate the correlated atoms. $\sigma= 2.78$~bohr and $\rho_{\text{th}}=0.015$~e/bohr$^{3}$ were used in the calculations. The results for bulk Ni (FM), NiO (AFM), Fe (FM), and FeO (AFM) are also shown for comparison. Fe-full coverage(2) is a $2 \times 2$ Fe surface with full oxygen coverage and one extra oxygen in the subsurface and (*) indicates the Fe atom on the surface which also has one oxygen neighbor in the subsurface.}
    \label{Ni-Fe-surface:Ta}
    \begin{ruledtabular}
    \begin{tabular}{lcc}
System&$\tilde{g}$&$\tilde{s}$\\
\hline
Ni (Ni@surface)         &1.52&0.51\\
Ni (Ni@subsurface)         &1.54&0.49\\
Ni (Ni@middle)           &1.52&0.48\\
\hline
Ni-full coverage (Ni@surface)     &\textbf{1.71}&\textbf{0.56}\\
Ni-full coverage (Ni@subsurface)   &1.56&0.49\\
Ni-full coverage (Ni@middle)      &1.54&0.49\\
\hline
Ni(bulk)                             &1.52&0.48\\
NiO(bulk)                            &\textbf{1.81}&\textbf{0.65}\\
\hline
Fe (Fe@surface)  &1.40&0.47\\
Fe (Fe@subsurface) &1.42&0.49\\
Fe (Fe@middle)   &1.38&0.47\\
\hline
\hline
Fe-full coverage (Fe@surface)  &\textbf{1.60}&\textbf{0.55}\\
Fe-full coverage (Fe@subsurface) &1.48&051\\
Fe-full coverage (Fe@middle)   &1.39&0.47\\
\hline
Fe-full coverage(2) (Fe@surface*)  &\textbf{1.75}&\textbf{0.56}\\
Fe-full coverage(2) (Fe@surface)  &\textbf{1.61}&\textbf{0.56}\\
Fe-full coverage(2) (Fe@subsurface) &1.52&0.53\\
Fe-full coverage(2) (Fe@middle)   &1.41&0.49\\
\hline 
Fe-half coverage (Fe@surface)   &1.55&0.52\\
Fe-half coverage (Fe@subsurface(near O))  &1.47&0.50\\
Fe-half coverage (Fe@subsurface(far O))   &1.46&0.50\\
Fe-half coverage (Fe@middle)          &1.39&0.47\\
\hline
Fe-1/9 coverage (Fe@surface(near O))   &1.45&0.49\\
Fe-1/9 coverage (Fe@surface (far O))    &1.40&0.47\\
Fe-1/9 coverage (Fe@subsurface(near O))  &1.46&0.50\\
Fe-1/9 coverage (Fe@subsurface(far O))   &1.42&0.49\\
Fe-1/9 coverage (Fe@middle)          &1.38&0.47\\
\hline
Fe(bulk)                                      &1.37&0.47\\
FeO(bulk)                                     &\textbf{1.73}&\textbf{0.64}\\
\end{tabular}
\end{ruledtabular}
\end{table}

\begin{figure}
    \centering
     \includegraphics[width=\columnwidth]{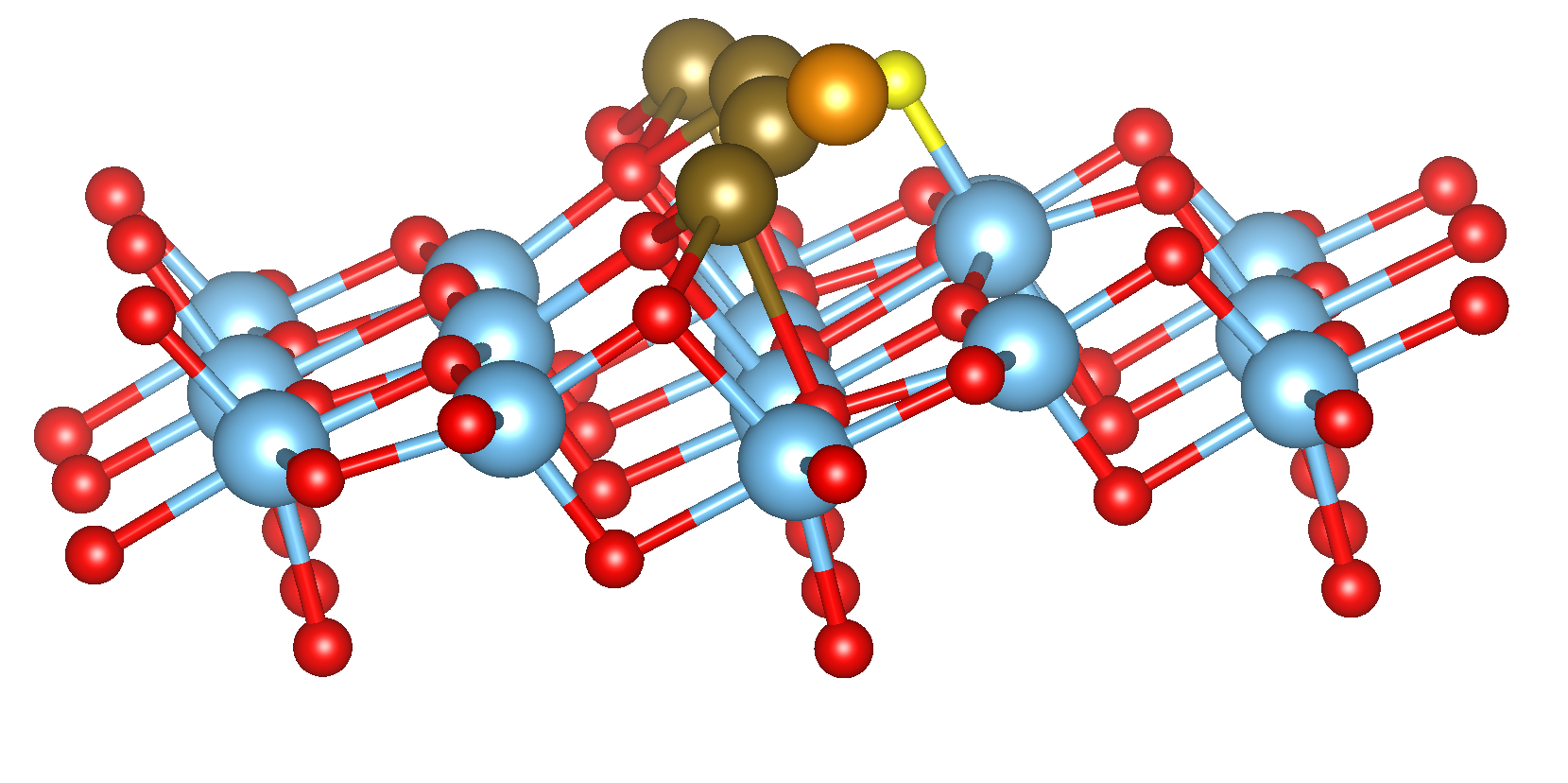}
     \caption{Side view of a Cu$_{5}$O cluster adsorbed on the anatase TiO$_{2}$(101) surface. The color coding of the atoms is as follows: red (surface O), yellow (O bonded to Cu), blue (Ti), brown (correlated Cu), orange (non-correlated Cu).}
    \label{TiO2_Cu5O}
\end{figure}

We also calculated $\tilde{g}$ and $\tilde{s}$ on the Cu atoms in the more complex system shown in Fig.~\ref{TiO2_Cu5O}. It consists of a nonmagnetic Cu$_{5}$O cluster adsorbed on the anatase TiO$_{2}$(101) ($3 \times 2$) surface. This and similar systems studied in Ref.~\cite{SchubertAdM21} constitute very irregular surfaces with Cu atoms ranging from neutral to +2-charged. The results obtained with $\sigma=2.78$~bohr are shown in Table~\ref{TiO2-Cu5O-Table}. As discussed in Sec.~\ref{bulk}, the correlation estimators are mainly determined by the environment of the TM atom and the results show that the correlated Cu atoms (Cu1$-$Cu4) are those which are bonded to O atoms of the surface or cluster and have values above 1.7 for $\tilde{g}$ and at $\sim0.6$ for $\tilde{s}$. They are thus larger than for the non-correlated Cu5 atom ($\tilde{g}=1.59$ and $\tilde{s}=0.48$) that is bonded only to another Cu atom. The results for the bulk solids Cu (NM), CuO (AFM), and Cu$_{2}$O (NM) are also shown in Table~\ref{TiO2-Cu5O-Table}, and we can see that the values for bulk Cu are similar to the values for the Cu5 atom. The values of $\tilde{g}$ for CuO and Cu$_{2}$O are larger and smaller than for Cu1$-$Cu4, respectively. For $\tilde{s}$, the values are larger in both CuO and Cu$_{2}$O compared to Cu1$-$Cu4.

Obviously, also the oxidation state depends on the environment and it is interesting to calculate the Bader charges \cite{doi:10.1021/ar00109a003} of the Cu atoms as an additional quantity to compare with. The calculation of the Bader charge is based on the gradient vector field of $\rho$, and the surfaces of zero flux define the atoms and therefore their charge (nuclear charge minus integrated $\rho$).
As shown in Table~\ref{TiO2-Cu5O-Table} the Bader charge is almost zero for the non-correlated Cu5, while it amounts to 0.60 for Cu1 and Cu2 and 0.41 for Cu3 and Cu4, corresponding to Cu$^+$ ions, which were estimated as correlated atoms. 

\begin{table}
    \caption{$\tilde{g}$, $\tilde{s}$, and Bader charge of the five Cu atoms of the Cu$_{5}$O cluster adsorbed on the anatase TiO$_{2}$(101) surface. The values in bold indicate the correlated atoms. $\sigma$ = 2.78~bohr and $\rho_{\text{th}}=0.015$~e/bohr$^{3}$ were used in the calculations. The results for Cu in bulk Cu (NM), CuO (AFM), and Cu$_{2}$O (NM) are also shown for comparison.}
    \label{TiO2-Cu5O-Table}
    \begin{ruledtabular}
    \begin{tabular}{lccc}
System &$\tilde{g}$&$\tilde{s}$& Bader charge\\
\hline
TiO$_{2}$-Cu$_{5}$O(Cu1)         &\textbf{1.74}&\textbf{0.58}&0.60\\
TiO$_{2}$-Cu$_{5}$O(Cu2)         &\textbf{1.74}&\textbf{0.58}&0.60\\
TiO$_{2}$-Cu$_{5}$O(Cu3)         &\textbf{1.72}&\textbf{0.59}&0.41\\
TiO$_{2}$-Cu$_{5}$O(Cu4)         &\textbf{1.72}&\textbf{0.59}&0.41\\
TiO$_{2}$-Cu$_{5}$O(Cu5)         &1.59&0.48&0.07\\
\hline
Cu(bulk)             &1.51&0.48&0\\ 
CuO(bulk)            &\textbf{1.80}&\textbf{0.77}&1.07\\
Cu$_{2}$O(bulk)           &\textbf{1.64}&\textbf{0.67}&0.55\\
    \end{tabular}
        \end{ruledtabular}
\end{table}

Thus, overall there are very clear trends in the values of the correlation estimators on the TM atoms also in complicated systems. 
Thus, this demonstrates that $\tilde{g}$ or $\tilde{s}$ can be efficiently used to estimate the correlation on TM atoms in complex systems like interfaces or surfaces.

\section{\label{summary}Summary}

In this work, we have shown that the value of the correlation estimators $\tilde{g}$ and $\tilde{s}$ at the nucleus of a TM atom, which are local averages of density-dependent quantities around the corresponding atom, can be used to estimate the strength of correlation of the TM atom. In bulk solids, where we usually know from experience in which systems the TM atoms are correlated, there is a very clear difference in the values of the correlation estimators between correlated (e.g., in oxides) and non-correlated (e.g., in pure metals) TM atoms. In more complicated systems, like at interfaces or surfaces, it may be unclear whether a certain TM atom is correlated or not, however we showed that our correlation estimators are very reliable in providing a very good hint on the correlation strength. We have demonstrated the power of the estimators for oxygen-covered TM surfaces, for Cu$_{5}$O clusters adsorbed on the TiO$_{2}$-anatase surface, and a Mn/MnO$_2$ interface. Thus, $\tilde{g}$ or $\tilde{s}$ could be used to determine for which atoms a Hubbard $U$ correction should be applied in a DFT+$U$ calculation. According to the results shown in this work, we would favour $\tilde{g}$ as a more reliable estimator. However, it does not seem possible to go to a more quantitative level and to find a relation between the estimators and a specific value of $U$ in general, although it might be possible to use $\tilde{g}$ or $\tilde{s}$ in systems having several TM atoms of the same type but in different environments as demonstrated for the Mn/MnO$_2$ interface or the Fe and Ni surfaces.

\begin{acknowledgements}
L.K. and P.B. acknowledge support by the TU-D doctoral college (TU Wien).
\end{acknowledgements}

\bibliography{references}

\begin{thebibliography}{43}%
\makeatletter
\providecommand \@ifxundefined [1]{%
 \@ifx{#1\undefined}
}%
\providecommand \@ifnum [1]{%
 \ifnum #1\expandafter \@firstoftwo
 \else \expandafter \@secondoftwo
 \fi
}%
\providecommand \@ifx [1]{%
 \ifx #1\expandafter \@firstoftwo
 \else \expandafter \@secondoftwo
 \fi
}%
\providecommand \natexlab [1]{#1}%
\providecommand \enquote  [1]{``#1''}%
\providecommand \bibnamefont  [1]{#1}%
\providecommand \bibfnamefont [1]{#1}%
\providecommand \citenamefont [1]{#1}%
\providecommand \href@noop [0]{\@secondoftwo}%
\providecommand \href [0]{\begingroup \@sanitize@url \@href}%
\providecommand \@href[1]{\@@startlink{#1}\@@href}%
\providecommand \@@href[1]{\endgroup#1\@@endlink}%
\providecommand \@sanitize@url [0]{\catcode `\\12\catcode `\$12\catcode
  `\&12\catcode `\#12\catcode `\^12\catcode `\_12\catcode `\%12\relax}%
\providecommand \@@startlink[1]{}%
\providecommand \@@endlink[0]{}%
\providecommand \url  [0]{\begingroup\@sanitize@url \@url }%
\providecommand \@url [1]{\endgroup\@href {#1}{\urlprefix }}%
\providecommand \urlprefix  [0]{URL }%
\providecommand \Eprint [0]{\href }%
\providecommand \doibase [0]{https://doi.org/}%
\providecommand \selectlanguage [0]{\@gobble}%
\providecommand \bibinfo  [0]{\@secondoftwo}%
\providecommand \bibfield  [0]{\@secondoftwo}%
\providecommand \translation [1]{[#1]}%
\providecommand \BibitemOpen [0]{}%
\providecommand \bibitemStop [0]{}%
\providecommand \bibitemNoStop [0]{.\EOS\space}%
\providecommand \EOS [0]{\spacefactor3000\relax}%
\providecommand \BibitemShut  [1]{\csname bibitem#1\endcsname}%
\let\auto@bib@innerbib\@empty
\bibitem [{\citenamefont {Hohenberg}\ and\ \citenamefont
  {Kohn}(1964)}]{HohenbergPR64}%
  \BibitemOpen
  \bibfield  {author} {\bibinfo {author} {\bibfnamefont {P.}~\bibnamefont
  {Hohenberg}}\ and\ \bibinfo {author} {\bibfnamefont {W.}~\bibnamefont
  {Kohn}},\ }\href@noop {} {\bibfield  {journal} {\bibinfo  {journal} {Phys.
  Rev.}\ }\textbf {\bibinfo {volume} {136}},\ \bibinfo {pages} {B864} (\bibinfo
  {year} {1964})}\BibitemShut {NoStop}%
\bibitem [{\citenamefont {Kohn}\ and\ \citenamefont {Sham}(1965)}]{KohnPR65}%
  \BibitemOpen
  \bibfield  {author} {\bibinfo {author} {\bibfnamefont {W.}~\bibnamefont
  {Kohn}}\ and\ \bibinfo {author} {\bibfnamefont {L.~J.}\ \bibnamefont
  {Sham}},\ }\href@noop {} {\bibfield  {journal} {\bibinfo  {journal} {Phys.
  Rev.}\ }\textbf {\bibinfo {volume} {140}},\ \bibinfo {pages} {A1133}
  (\bibinfo {year} {1965})}\BibitemShut {NoStop}%
\bibitem [{\citenamefont {Terakura}\ \emph {et~al.}(1984)\citenamefont
  {Terakura}, \citenamefont {Oguchi}, \citenamefont {Williams},\ and\
  \citenamefont {K\"{u}bler}}]{TerakuraPRB84}%
  \BibitemOpen
  \bibfield  {author} {\bibinfo {author} {\bibfnamefont {K.}~\bibnamefont
  {Terakura}}, \bibinfo {author} {\bibfnamefont {T.}~\bibnamefont {Oguchi}},
  \bibinfo {author} {\bibfnamefont {A.~R.}\ \bibnamefont {Williams}},\ and\
  \bibinfo {author} {\bibfnamefont {J.}~\bibnamefont {K\"{u}bler}},\
  }\href@noop {} {\bibfield  {journal} {\bibinfo  {journal} {Phys. Rev. B}\
  }\textbf {\bibinfo {volume} {30}},\ \bibinfo {pages} {4734} (\bibinfo {year}
  {1984})}\BibitemShut {NoStop}%
\bibitem [{\citenamefont {Bredow}\ and\ \citenamefont
  {Gerson}(2000)}]{BredowPRB00}%
  \BibitemOpen
  \bibfield  {author} {\bibinfo {author} {\bibfnamefont {T.}~\bibnamefont
  {Bredow}}\ and\ \bibinfo {author} {\bibfnamefont {A.~R.}\ \bibnamefont
  {Gerson}},\ }\href@noop {} {\bibfield  {journal} {\bibinfo  {journal} {Phys.
  Rev. B}\ }\textbf {\bibinfo {volume} {61}},\ \bibinfo {pages} {5194}
  (\bibinfo {year} {2000})}\BibitemShut {NoStop}%
\bibitem [{\citenamefont {Perry}\ \emph {et~al.}(2001)\citenamefont {Perry},
  \citenamefont {Tahir-Kheli},\ and\ \citenamefont {Goddard}}]{PerryPRB01}%
  \BibitemOpen
  \bibfield  {author} {\bibinfo {author} {\bibfnamefont {J.~K.}\ \bibnamefont
  {Perry}}, \bibinfo {author} {\bibfnamefont {J.}~\bibnamefont {Tahir-Kheli}},\
  and\ \bibinfo {author} {\bibfnamefont {W.~A.}\ \bibnamefont {Goddard},
  \bibfnamefont {III}},\ }\href@noop {} {\bibfield  {journal} {\bibinfo
  {journal} {Phys. Rev. B}\ }\textbf {\bibinfo {volume} {63}},\ \bibinfo
  {pages} {144510} (\bibinfo {year} {2001})}\BibitemShut {NoStop}%
\bibitem [{\citenamefont {Muscat}\ \emph {et~al.}(2001)\citenamefont {Muscat},
  \citenamefont {Wander},\ and\ \citenamefont {Harrison}}]{MuscatCPL01}%
  \BibitemOpen
  \bibfield  {author} {\bibinfo {author} {\bibfnamefont {J.}~\bibnamefont
  {Muscat}}, \bibinfo {author} {\bibfnamefont {A.}~\bibnamefont {Wander}},\
  and\ \bibinfo {author} {\bibfnamefont {N.~M.}\ \bibnamefont {Harrison}},\
  }\href@noop {} {\bibfield  {journal} {\bibinfo  {journal} {Chem. Phys.
  Lett.}\ }\textbf {\bibinfo {volume} {342}},\ \bibinfo {pages} {397} (\bibinfo
  {year} {2001})}\BibitemShut {NoStop}%
\bibitem [{\citenamefont {Heyd}\ \emph {et~al.}(2003)\citenamefont {Heyd},
  \citenamefont {Scuseria},\ and\ \citenamefont {Ernzerhof}}]{HeydJCP03}%
  \BibitemOpen
  \bibfield  {author} {\bibinfo {author} {\bibfnamefont {J.}~\bibnamefont
  {Heyd}}, \bibinfo {author} {\bibfnamefont {G.~E.}\ \bibnamefont {Scuseria}},\
  and\ \bibinfo {author} {\bibfnamefont {M.}~\bibnamefont {Ernzerhof}},\
  }\href@noop {} {\bibfield  {journal} {\bibinfo  {journal} {J. Chem. Phys.}\
  }\textbf {\bibinfo {volume} {118}},\ \bibinfo {pages} {8207} (\bibinfo {year}
  {2003})},\ \bibinfo {note} {\textbf{124}, 219906 (2006)}\BibitemShut
  {NoStop}%
\bibitem [{\citenamefont {Metzner}\ and\ \citenamefont
  {Vollhardt}(1989)}]{MetznerPRL89}%
  \BibitemOpen
  \bibfield  {author} {\bibinfo {author} {\bibfnamefont {W.}~\bibnamefont
  {Metzner}}\ and\ \bibinfo {author} {\bibfnamefont {D.}~\bibnamefont
  {Vollhardt}},\ }\href@noop {} {\bibfield  {journal} {\bibinfo  {journal}
  {Phys. Rev. Lett.}\ }\textbf {\bibinfo {volume} {62}},\ \bibinfo {pages}
  {324} (\bibinfo {year} {1989})},\ \bibinfo {note} {\textbf{62}, 1066
  (1989)}\BibitemShut {NoStop}%
\bibitem [{\citenamefont {Kotliar}\ \emph {et~al.}(2006)\citenamefont
  {Kotliar}, \citenamefont {Savrasov}, \citenamefont {Haule}, \citenamefont
  {Oudovenko}, \citenamefont {Parcollet},\ and\ \citenamefont
  {Marianetti}}]{KotliarRMP06}%
  \BibitemOpen
  \bibfield  {author} {\bibinfo {author} {\bibfnamefont {G.}~\bibnamefont
  {Kotliar}}, \bibinfo {author} {\bibfnamefont {S.~Y.}\ \bibnamefont
  {Savrasov}}, \bibinfo {author} {\bibfnamefont {K.}~\bibnamefont {Haule}},
  \bibinfo {author} {\bibfnamefont {V.~S.}\ \bibnamefont {Oudovenko}}, \bibinfo
  {author} {\bibfnamefont {O.}~\bibnamefont {Parcollet}},\ and\ \bibinfo
  {author} {\bibfnamefont {C.~A.}\ \bibnamefont {Marianetti}},\ }\href@noop {}
  {\bibfield  {journal} {\bibinfo  {journal} {Rev. Mod. Phys.}\ }\textbf
  {\bibinfo {volume} {78}},\ \bibinfo {pages} {865} (\bibinfo {year}
  {2006})}\BibitemShut {NoStop}%
\bibitem [{\citenamefont {Held}(2007)}]{HeldAP07}%
  \BibitemOpen
  \bibfield  {author} {\bibinfo {author} {\bibfnamefont {K.}~\bibnamefont
  {Held}},\ }\href@noop {} {\bibfield  {journal} {\bibinfo  {journal} {Adv.
  Phys.}\ }\textbf {\bibinfo {volume} {56}},\ \bibinfo {pages} {829} (\bibinfo
  {year} {2007})}\BibitemShut {NoStop}%
\bibitem [{\citenamefont {Sharma}\ \emph {et~al.}(2013)\citenamefont {Sharma},
  \citenamefont {Dewhurst}, \citenamefont {Shallcross},\ and\ \citenamefont
  {Gross}}]{SharmaPRL13}%
  \BibitemOpen
  \bibfield  {author} {\bibinfo {author} {\bibfnamefont {S.}~\bibnamefont
  {Sharma}}, \bibinfo {author} {\bibfnamefont {J.~K.}\ \bibnamefont
  {Dewhurst}}, \bibinfo {author} {\bibfnamefont {S.}~\bibnamefont
  {Shallcross}},\ and\ \bibinfo {author} {\bibfnamefont {E.~K.~U.}\
  \bibnamefont {Gross}},\ }\href@noop {} {\bibfield  {journal} {\bibinfo
  {journal} {Phys. Rev. Lett.}\ }\textbf {\bibinfo {volume} {110}},\ \bibinfo
  {pages} {116403} (\bibinfo {year} {2013})}\BibitemShut {NoStop}%
\bibitem [{\citenamefont {Di~Sabatino}\ \emph {et~al.}(2015)\citenamefont
  {Di~Sabatino}, \citenamefont {Berger}, \citenamefont {Reining},\ and\
  \citenamefont {Romaniello}}]{DiSabatinoJCP15}%
  \BibitemOpen
  \bibfield  {author} {\bibinfo {author} {\bibfnamefont {S.}~\bibnamefont
  {Di~Sabatino}}, \bibinfo {author} {\bibfnamefont {J.~A.}\ \bibnamefont
  {Berger}}, \bibinfo {author} {\bibfnamefont {L.}~\bibnamefont {Reining}},\
  and\ \bibinfo {author} {\bibfnamefont {P.}~\bibnamefont {Romaniello}},\
  }\href@noop {} {\bibfield  {journal} {\bibinfo  {journal} {J. Chem. Pyhs.}\
  }\textbf {\bibinfo {volume} {143}},\ \bibinfo {pages} {024108} (\bibinfo
  {year} {2015})}\BibitemShut {NoStop}%
\bibitem [{\citenamefont {Tran}\ and\ \citenamefont {Blaha}(2009)}]{TranPRL09}%
  \BibitemOpen
  \bibfield  {author} {\bibinfo {author} {\bibfnamefont {F.}~\bibnamefont
  {Tran}}\ and\ \bibinfo {author} {\bibfnamefont {P.}~\bibnamefont {Blaha}},\
  }\href@noop {} {\bibfield  {journal} {\bibinfo  {journal} {Phys. Rev. Lett.}\
  }\textbf {\bibinfo {volume} {102}},\ \bibinfo {pages} {226401} (\bibinfo
  {year} {2009})}\BibitemShut {NoStop}%
\bibitem [{\citenamefont {Tran}\ and\ \citenamefont
  {Blaha}(2017)}]{TranJPCA17}%
  \BibitemOpen
  \bibfield  {author} {\bibinfo {author} {\bibfnamefont {F.}~\bibnamefont
  {Tran}}\ and\ \bibinfo {author} {\bibfnamefont {P.}~\bibnamefont {Blaha}},\
  }\href@noop {} {\bibfield  {journal} {\bibinfo  {journal} {J. Phys. Chem. A}\
  }\textbf {\bibinfo {volume} {121}},\ \bibinfo {pages} {3318} (\bibinfo {year}
  {2017})}\BibitemShut {NoStop}%
\bibitem [{\citenamefont {Anisimov}\ \emph {et~al.}(1993)\citenamefont
  {Anisimov}, \citenamefont {Solovyev}, \citenamefont {Korotin}, \citenamefont
  {Czy\ifmmode~\dot{z}\else \.{z}\fi{}yk},\ and\ \citenamefont
  {Sawatzky}}]{AnisimovPRB93}%
  \BibitemOpen
  \bibfield  {author} {\bibinfo {author} {\bibfnamefont {V.~I.}\ \bibnamefont
  {Anisimov}}, \bibinfo {author} {\bibfnamefont {I.~V.}\ \bibnamefont
  {Solovyev}}, \bibinfo {author} {\bibfnamefont {M.~A.}\ \bibnamefont
  {Korotin}}, \bibinfo {author} {\bibfnamefont {M.~T.}\ \bibnamefont
  {Czy\ifmmode~\dot{z}\else \.{z}\fi{}yk}},\ and\ \bibinfo {author}
  {\bibfnamefont {G.~A.}\ \bibnamefont {Sawatzky}},\ }\href@noop {} {\bibfield
  {journal} {\bibinfo  {journal} {Phys. Rev. B}\ }\textbf {\bibinfo {volume}
  {48}},\ \bibinfo {pages} {16929} (\bibinfo {year} {1993})}\BibitemShut
  {NoStop}%
\bibitem [{\citenamefont {Ylvisaker}\ \emph {et~al.}(2009)\citenamefont
  {Ylvisaker}, \citenamefont {Pickett},\ and\ \citenamefont
  {Koepernik}}]{YlvisakerPRB09}%
  \BibitemOpen
  \bibfield  {author} {\bibinfo {author} {\bibfnamefont {E.~R.}\ \bibnamefont
  {Ylvisaker}}, \bibinfo {author} {\bibfnamefont {W.~E.}\ \bibnamefont
  {Pickett}},\ and\ \bibinfo {author} {\bibfnamefont {K.}~\bibnamefont
  {Koepernik}},\ }\href@noop {} {\bibfield  {journal} {\bibinfo  {journal}
  {Phys. Rev. B}\ }\textbf {\bibinfo {volume} {79}},\ \bibinfo {pages} {035103}
  (\bibinfo {year} {2009})}\BibitemShut {NoStop}%
\bibitem [{\citenamefont {Himmetoglu}\ \emph {et~al.}(2014)\citenamefont
  {Himmetoglu}, \citenamefont {Floris}, \citenamefont {de~Gironcoli},\ and\
  \citenamefont {Cococcioni}}]{HimmetogluIJQC14}%
  \BibitemOpen
  \bibfield  {author} {\bibinfo {author} {\bibfnamefont {B.}~\bibnamefont
  {Himmetoglu}}, \bibinfo {author} {\bibfnamefont {A.}~\bibnamefont {Floris}},
  \bibinfo {author} {\bibfnamefont {S.}~\bibnamefont {de~Gironcoli}},\ and\
  \bibinfo {author} {\bibfnamefont {M.}~\bibnamefont {Cococcioni}},\
  }\href@noop {} {\bibfield  {journal} {\bibinfo  {journal} {Int. J. Quantum
  Chem.}\ }\textbf {\bibinfo {volume} {114}},\ \bibinfo {pages} {14} (\bibinfo
  {year} {2014})}\BibitemShut {NoStop}%
\bibitem [{\citenamefont {Dederichs}\ \emph {et~al.}(1984)\citenamefont
  {Dederichs}, \citenamefont {Bl\"ugel}, \citenamefont {Zeller},\ and\
  \citenamefont {Akai}}]{DederichsPRL84}%
  \BibitemOpen
  \bibfield  {author} {\bibinfo {author} {\bibfnamefont {P.~H.}\ \bibnamefont
  {Dederichs}}, \bibinfo {author} {\bibfnamefont {S.}~\bibnamefont {Bl\"ugel}},
  \bibinfo {author} {\bibfnamefont {R.}~\bibnamefont {Zeller}},\ and\ \bibinfo
  {author} {\bibfnamefont {H.}~\bibnamefont {Akai}},\ }\href@noop {} {\bibfield
   {journal} {\bibinfo  {journal} {Phys. Rev. Lett.}\ }\textbf {\bibinfo
  {volume} {53}},\ \bibinfo {pages} {2512} (\bibinfo {year}
  {1984})}\BibitemShut {NoStop}%
\bibitem [{\citenamefont {Hybertsen}\ \emph {et~al.}(1989)\citenamefont
  {Hybertsen}, \citenamefont {Schl\"uter},\ and\ \citenamefont
  {Christensen}}]{HybertsenPRB89}%
  \BibitemOpen
  \bibfield  {author} {\bibinfo {author} {\bibfnamefont {M.~S.}\ \bibnamefont
  {Hybertsen}}, \bibinfo {author} {\bibfnamefont {M.}~\bibnamefont
  {Schl\"uter}},\ and\ \bibinfo {author} {\bibfnamefont {N.~E.}\ \bibnamefont
  {Christensen}},\ }\href@noop {} {\bibfield  {journal} {\bibinfo  {journal}
  {Phys. Rev. B}\ }\textbf {\bibinfo {volume} {39}},\ \bibinfo {pages} {9028}
  (\bibinfo {year} {1989})}\BibitemShut {NoStop}%
\bibitem [{\citenamefont {Madsen}\ and\ \citenamefont
  {Nov\'{a}k}(2005)}]{MadsenEPL05}%
  \BibitemOpen
  \bibfield  {author} {\bibinfo {author} {\bibfnamefont {G.~K.~H.}\
  \bibnamefont {Madsen}}\ and\ \bibinfo {author} {\bibfnamefont
  {P.}~\bibnamefont {Nov\'{a}k}},\ }\href@noop {} {\bibfield  {journal}
  {\bibinfo  {journal} {Europhys. Lett.}\ }\textbf {\bibinfo {volume} {69}},\
  \bibinfo {pages} {777} (\bibinfo {year} {2005})}\BibitemShut {NoStop}%
\bibitem [{\citenamefont {Vaguier}\ \emph {et~al.}(2012)\citenamefont
  {Vaguier}, \citenamefont {Jiang},\ and\ \citenamefont
  {Biermann}}]{VaugierPRB2012}%
  \BibitemOpen
  \bibfield  {author} {\bibinfo {author} {\bibfnamefont {L.}~\bibnamefont
  {Vaguier}}, \bibinfo {author} {\bibfnamefont {H.}~\bibnamefont {Jiang}},\
  and\ \bibinfo {author} {\bibfnamefont {S.}~\bibnamefont {Biermann}},\
  }\href@noop {} {\bibfield  {journal} {\bibinfo  {journal} {Phys. Rev. B.}\
  }\textbf {\bibinfo {volume} {86}},\ \bibinfo {pages} {165105} (\bibinfo
  {year} {2012})}\BibitemShut {NoStop}%
\bibitem [{\citenamefont {Springer}\ and\ \citenamefont
  {Aryasetiawan}(1998)}]{SpringerPRB98}%
  \BibitemOpen
  \bibfield  {author} {\bibinfo {author} {\bibfnamefont {M.}~\bibnamefont
  {Springer}}\ and\ \bibinfo {author} {\bibfnamefont {F.}~\bibnamefont
  {Aryasetiawan}},\ }\href@noop {} {\bibfield  {journal} {\bibinfo  {journal}
  {Phys. Rev. B}\ }\textbf {\bibinfo {volume} {57}},\ \bibinfo {pages} {4364}
  (\bibinfo {year} {1998})}\BibitemShut {NoStop}%
\bibitem [{\citenamefont {Aryasetiawan}\ \emph {et~al.}(2004)\citenamefont
  {Aryasetiawan}, \citenamefont {Imada}, \citenamefont {Georges}, \citenamefont
  {Kotliar}, \citenamefont {Biermann},\ and\ \citenamefont
  {Lichtenstein}}]{AryasetiawanPRB2004}%
  \BibitemOpen
  \bibfield  {author} {\bibinfo {author} {\bibfnamefont {F.}~\bibnamefont
  {Aryasetiawan}}, \bibinfo {author} {\bibfnamefont {M.}~\bibnamefont {Imada}},
  \bibinfo {author} {\bibfnamefont {A.}~\bibnamefont {Georges}}, \bibinfo
  {author} {\bibfnamefont {G.}~\bibnamefont {Kotliar}}, \bibinfo {author}
  {\bibfnamefont {S.}~\bibnamefont {Biermann}},\ and\ \bibinfo {author}
  {\bibfnamefont {A.~I.}\ \bibnamefont {Lichtenstein}},\ }\href@noop {}
  {\bibfield  {journal} {\bibinfo  {journal} {Phys. Rev. B}\ }\textbf {\bibinfo
  {volume} {70}},\ \bibinfo {pages} {195104} (\bibinfo {year}
  {2004})}\BibitemShut {NoStop}%
\bibitem [{\citenamefont {Aryasetiawan}\ \emph {et~al.}(2006)\citenamefont
  {Aryasetiawan}, \citenamefont {Karlsson}, \citenamefont {Jepsen},\ and\
  \citenamefont {Sch\"onberger}}]{AryasetiawanPRB2006}%
  \BibitemOpen
  \bibfield  {author} {\bibinfo {author} {\bibfnamefont {F.}~\bibnamefont
  {Aryasetiawan}}, \bibinfo {author} {\bibfnamefont {K.}~\bibnamefont
  {Karlsson}}, \bibinfo {author} {\bibfnamefont {O.}~\bibnamefont {Jepsen}},\
  and\ \bibinfo {author} {\bibfnamefont {U.}~\bibnamefont {Sch\"onberger}},\
  }\href@noop {} {\bibfield  {journal} {\bibinfo  {journal} {Phys. Rev. B}\
  }\textbf {\bibinfo {volume} {74}},\ \bibinfo {pages} {125106} (\bibinfo
  {year} {2006})}\BibitemShut {NoStop}%
\bibitem [{\citenamefont {\ifmmode \mbox{\c{S}}\else \c{S}\fi{}a\ifmmode
  \mbox{\c{s}}\else \c{s}\fi{}\ifmmode \imath \else \i
  \fi{}o\ifmmode~\breve{g}\else \u{g}\fi{}lu}\ \emph
  {et~al.}(2012)\citenamefont {\ifmmode \mbox{\c{S}}\else \c{S}\fi{}a\ifmmode
  \mbox{\c{s}}\else \c{s}\fi{}\ifmmode \imath \else \i
  \fi{}o\ifmmode~\breve{g}\else \u{g}\fi{}lu}, \citenamefont {Friedrich},\ and\
  \citenamefont {Bl\"ugel}}]{SasiogluPRB2012}%
  \BibitemOpen
  \bibfield  {author} {\bibinfo {author} {\bibfnamefont {E.}~\bibnamefont
  {\ifmmode \mbox{\c{S}}\else \c{S}\fi{}a\ifmmode \mbox{\c{s}}\else
  \c{s}\fi{}\ifmmode \imath \else \i \fi{}o\ifmmode~\breve{g}\else
  \u{g}\fi{}lu}}, \bibinfo {author} {\bibfnamefont {C.}~\bibnamefont
  {Friedrich}},\ and\ \bibinfo {author} {\bibfnamefont {S.}~\bibnamefont
  {Bl\"ugel}},\ }\href@noop {} {\bibfield  {journal} {\bibinfo  {journal}
  {Phys. Rev. Lett.}\ }\textbf {\bibinfo {volume} {109}},\ \bibinfo {pages}
  {146401} (\bibinfo {year} {2012})}\BibitemShut {NoStop}%
\bibitem [{\citenamefont {Cococcioni}\ and\ \citenamefont
  {de~Gironcoli}(2005)}]{CococcioniPRB2005}%
  \BibitemOpen
  \bibfield  {author} {\bibinfo {author} {\bibfnamefont {M.}~\bibnamefont
  {Cococcioni}}\ and\ \bibinfo {author} {\bibfnamefont {S.}~\bibnamefont
  {de~Gironcoli}},\ }\href@noop {} {\bibfield  {journal} {\bibinfo  {journal}
  {Phys. Rev. B}\ }\textbf {\bibinfo {volume} {71}},\ \bibinfo {pages} {035105}
  (\bibinfo {year} {2005})}\BibitemShut {NoStop}%
\bibitem [{\citenamefont {Pickett}\ \emph {et~al.}(1998)\citenamefont
  {Pickett}, \citenamefont {Erwin},\ and\ \citenamefont
  {Ethridge}}]{PickettPRB98}%
  \BibitemOpen
  \bibfield  {author} {\bibinfo {author} {\bibfnamefont {W.~E.}\ \bibnamefont
  {Pickett}}, \bibinfo {author} {\bibfnamefont {S.~C.}\ \bibnamefont {Erwin}},\
  and\ \bibinfo {author} {\bibfnamefont {E.~C.}\ \bibnamefont {Ethridge}},\
  }\href@noop {} {\bibfield  {journal} {\bibinfo  {journal} {Phys. Rev. B}\
  }\textbf {\bibinfo {volume} {58}},\ \bibinfo {pages} {1201} (\bibinfo {year}
  {1998})}\BibitemShut {NoStop}%
\bibitem [{\citenamefont {Wang}\ and\ \citenamefont {Jiang}(2019)}]{WangJCP19}%
  \BibitemOpen
  \bibfield  {author} {\bibinfo {author} {\bibfnamefont {Y.-C.}\ \bibnamefont
  {Wang}}\ and\ \bibinfo {author} {\bibfnamefont {H.}~\bibnamefont {Jiang}},\
  }\href@noop {} {\bibfield  {journal} {\bibinfo  {journal} {J. Chem. Phys.}\
  }\textbf {\bibinfo {volume} {150}},\ \bibinfo {pages} {154116} (\bibinfo
  {year} {2019})}\BibitemShut {NoStop}%
\bibitem [{\citenamefont {Krukau}\ \emph {et~al.}(2006)\citenamefont {Krukau},
  \citenamefont {Vydrov}, \citenamefont {Izmaylov},\ and\ \citenamefont
  {Scuseria}}]{KrukauJCP06}%
  \BibitemOpen
  \bibfield  {author} {\bibinfo {author} {\bibfnamefont {A.~V.}\ \bibnamefont
  {Krukau}}, \bibinfo {author} {\bibfnamefont {O.~A.}\ \bibnamefont {Vydrov}},
  \bibinfo {author} {\bibfnamefont {A.~F.}\ \bibnamefont {Izmaylov}},\ and\
  \bibinfo {author} {\bibfnamefont {G.~E.}\ \bibnamefont {Scuseria}},\
  }\href@noop {} {\bibfield  {journal} {\bibinfo  {journal} {J. Chem. Phys.}\
  }\textbf {\bibinfo {volume} {125}},\ \bibinfo {pages} {224106} (\bibinfo
  {year} {2006})}\BibitemShut {NoStop}%
\bibitem [{\citenamefont {Tran}\ \emph {et~al.}(2019)\citenamefont {Tran},
  \citenamefont {Doumont}, \citenamefont {Kalantari}, \citenamefont {Huran},
  \citenamefont {Marques},\ and\ \citenamefont {Blaha}}]{TranJAP19}%
  \BibitemOpen
  \bibfield  {author} {\bibinfo {author} {\bibfnamefont {F.}~\bibnamefont
  {Tran}}, \bibinfo {author} {\bibfnamefont {J.}~\bibnamefont {Doumont}},
  \bibinfo {author} {\bibfnamefont {L.}~\bibnamefont {Kalantari}}, \bibinfo
  {author} {\bibfnamefont {A.~W.}\ \bibnamefont {Huran}}, \bibinfo {author}
  {\bibfnamefont {M.~A.~L.}\ \bibnamefont {Marques}},\ and\ \bibinfo {author}
  {\bibfnamefont {P.}~\bibnamefont {Blaha}},\ }\href@noop {} {\bibfield
  {journal} {\bibinfo  {journal} {J. Appl. Phys.}\ }\textbf {\bibinfo {volume}
  {126}},\ \bibinfo {pages} {110902} (\bibinfo {year} {2019})}\BibitemShut
  {NoStop}%
\bibitem [{\citenamefont {Marques}\ \emph {et~al.}(2011)\citenamefont
  {Marques}, \citenamefont {Vidal}, \citenamefont {Oliveira}, \citenamefont
  {Reining},\ and\ \citenamefont {Botti}}]{MarquesPRB11}%
  \BibitemOpen
  \bibfield  {author} {\bibinfo {author} {\bibfnamefont {M.~A.~L.}\
  \bibnamefont {Marques}}, \bibinfo {author} {\bibfnamefont {J.}~\bibnamefont
  {Vidal}}, \bibinfo {author} {\bibfnamefont {M.~J.~T.}\ \bibnamefont
  {Oliveira}}, \bibinfo {author} {\bibfnamefont {L.}~\bibnamefont {Reining}},\
  and\ \bibinfo {author} {\bibfnamefont {S.}~\bibnamefont {Botti}},\
  }\href@noop {} {\bibfield  {journal} {\bibinfo  {journal} {Phys. Rev. B}\
  }\textbf {\bibinfo {volume} {83}},\ \bibinfo {pages} {035119} (\bibinfo
  {year} {2011})}\BibitemShut {NoStop}%
\bibitem [{\citenamefont {Rauch}\ \emph
  {et~al.}(2020{\natexlab{a}})\citenamefont {Rauch}, \citenamefont {Marques},\
  and\ \citenamefont {Botti}}]{RauchJCTC2020}%
  \BibitemOpen
  \bibfield  {author} {\bibinfo {author} {\bibfnamefont {T.}~\bibnamefont
  {Rauch}}, \bibinfo {author} {\bibfnamefont {M.~A.~L.}\ \bibnamefont
  {Marques}},\ and\ \bibinfo {author} {\bibfnamefont {S.}~\bibnamefont
  {Botti}},\ }\href@noop {} {\bibfield  {journal} {\bibinfo  {journal} {J.
  Chem. Theory Comput.}\ }\textbf {\bibinfo {volume} {16}},\ \bibinfo {pages}
  {2654} (\bibinfo {year} {2020}{\natexlab{a}})}\BibitemShut {NoStop}%
\bibitem [{\citenamefont {Rauch}\ \emph
  {et~al.}(2020{\natexlab{b}})\citenamefont {Rauch}, \citenamefont {Marques},\
  and\ \citenamefont {Botti}}]{RauchPRB20}%
  \BibitemOpen
  \bibfield  {author} {\bibinfo {author} {\bibfnamefont {T.}~\bibnamefont
  {Rauch}}, \bibinfo {author} {\bibfnamefont {M.~A.~L.}\ \bibnamefont
  {Marques}},\ and\ \bibinfo {author} {\bibfnamefont {S.}~\bibnamefont
  {Botti}},\ }\href@noop {} {\bibfield  {journal} {\bibinfo  {journal} {Phys.
  Rev. B}\ }\textbf {\bibinfo {volume} {101}},\ \bibinfo {pages} {245163}
  (\bibinfo {year} {2020}{\natexlab{b}})},\ \bibinfo {note} {\textbf{102},
  119902(E) (2020)}\BibitemShut {NoStop}%
\bibitem [{\citenamefont {Perdew}\ \emph {et~al.}(1996)\citenamefont {Perdew},
  \citenamefont {Burke},\ and\ \citenamefont {Ernzerhof}}]{PerdewPRL96}%
  \BibitemOpen
  \bibfield  {author} {\bibinfo {author} {\bibfnamefont {J.~P.}\ \bibnamefont
  {Perdew}}, \bibinfo {author} {\bibfnamefont {K.}~\bibnamefont {Burke}},\ and\
  \bibinfo {author} {\bibfnamefont {M.}~\bibnamefont {Ernzerhof}},\ }\href@noop
  {} {\bibfield  {journal} {\bibinfo  {journal} {Phys. Rev. Lett.}\ }\textbf
  {\bibinfo {volume} {77}},\ \bibinfo {pages} {3865} (\bibinfo {year}
  {1996})},\ \bibinfo {note} {\textbf{78}, 1396(E) (1997)}\BibitemShut
  {NoStop}%
\bibitem [{\citenamefont {Blaha}\ \emph {et~al.}(2018)\citenamefont {Blaha},
  \citenamefont {Schwarz}, \citenamefont {Madsen}, \citenamefont {Kvasnicka},
  \citenamefont {Luitz}, \citenamefont {Laskowski}, \citenamefont {Tran},\ and\
  \citenamefont {Marks}}]{WIEN2k}%
  \BibitemOpen
  \bibfield  {author} {\bibinfo {author} {\bibfnamefont {P.}~\bibnamefont
  {Blaha}}, \bibinfo {author} {\bibfnamefont {K.}~\bibnamefont {Schwarz}},
  \bibinfo {author} {\bibfnamefont {G.~K.~H.}\ \bibnamefont {Madsen}}, \bibinfo
  {author} {\bibfnamefont {D.}~\bibnamefont {Kvasnicka}}, \bibinfo {author}
  {\bibfnamefont {J.}~\bibnamefont {Luitz}}, \bibinfo {author} {\bibfnamefont
  {R.}~\bibnamefont {Laskowski}}, \bibinfo {author} {\bibfnamefont
  {F.}~\bibnamefont {Tran}},\ and\ \bibinfo {author} {\bibfnamefont {L.~D.}\
  \bibnamefont {Marks}},\ }\href@noop {} {\emph {\bibinfo {title} {WIEN2k: An
  Augmented Plane Wave plus Local Orbitals Program for Calculating Crystal
  Properties}}}\ (\bibinfo  {publisher} {Vienna University of Technology},\
  \bibinfo {address} {Austria},\ \bibinfo {year} {2018})\BibitemShut {NoStop}%
\bibitem [{\citenamefont {Blaha}\ \emph {et~al.}(2020)\citenamefont {Blaha},
  \citenamefont {Schwarz}, \citenamefont {Tran}, \citenamefont {Laskowski},
  \citenamefont {Madsen},\ and\ \citenamefont {Marks}}]{BlahaJCP20}%
  \BibitemOpen
  \bibfield  {author} {\bibinfo {author} {\bibfnamefont {P.}~\bibnamefont
  {Blaha}}, \bibinfo {author} {\bibfnamefont {K.}~\bibnamefont {Schwarz}},
  \bibinfo {author} {\bibfnamefont {F.}~\bibnamefont {Tran}}, \bibinfo {author}
  {\bibfnamefont {R.}~\bibnamefont {Laskowski}}, \bibinfo {author}
  {\bibfnamefont {G.~K.~H.}\ \bibnamefont {Madsen}},\ and\ \bibinfo {author}
  {\bibfnamefont {L.~D.}\ \bibnamefont {Marks}},\ }\href@noop {} {\bibfield
  {journal} {\bibinfo  {journal} {J. Chem. Phys.}\ }\textbf {\bibinfo {volume}
  {152}},\ \bibinfo {pages} {074101} (\bibinfo {year} {2020})}\BibitemShut
  {NoStop}%
\bibitem [{\citenamefont {Singh}\ and\ \citenamefont
  {Nordstr{\"{o}}m}(2006)}]{Singh}%
  \BibitemOpen
  \bibfield  {author} {\bibinfo {author} {\bibfnamefont {D.~J.}\ \bibnamefont
  {Singh}}\ and\ \bibinfo {author} {\bibfnamefont {L.}~\bibnamefont
  {Nordstr{\"{o}}m}},\ }\href@noop {} {\emph {\bibinfo {title} {Planewaves,
  Pseudopotentials and the LAPW Method, 2nd ed.}}}\ (\bibinfo  {publisher}
  {Springer},\ \bibinfo {address} {Berlin},\ \bibinfo {year}
  {2006})\BibitemShut {NoStop}%
\bibitem [{\citenamefont {Karsai}\ \emph {et~al.}(2017)\citenamefont {Karsai},
  \citenamefont {Tran},\ and\ \citenamefont {Blaha}}]{KarsaiCPC17}%
  \BibitemOpen
  \bibfield  {author} {\bibinfo {author} {\bibfnamefont {F.}~\bibnamefont
  {Karsai}}, \bibinfo {author} {\bibfnamefont {F.}~\bibnamefont {Tran}},\ and\
  \bibinfo {author} {\bibfnamefont {P.}~\bibnamefont {Blaha}},\ }\href@noop {}
  {\bibfield  {journal} {\bibinfo  {journal} {Comput. Phys. Commun.}\ }\textbf
  {\bibinfo {volume} {220}},\ \bibinfo {pages} {230} (\bibinfo {year}
  {2017})}\BibitemShut {NoStop}%
\bibitem [{\citenamefont {Singh}(1993)}]{SinghPRB93}%
  \BibitemOpen
  \bibfield  {author} {\bibinfo {author} {\bibfnamefont {D.~J.}\ \bibnamefont
  {Singh}},\ }\href@noop {} {\bibfield  {journal} {\bibinfo  {journal} {Phys.
  Rev. B}\ }\textbf {\bibinfo {volume} {48}},\ \bibinfo {pages} {14099}
  (\bibinfo {year} {1993})}\BibitemShut {NoStop}%
\bibitem [{\citenamefont {Jiang}(2013)}]{JiangJCP13}%
  \BibitemOpen
  \bibfield  {author} {\bibinfo {author} {\bibfnamefont {H.}~\bibnamefont
  {Jiang}},\ }\href@noop {} {\bibfield  {journal} {\bibinfo  {journal} {J.
  Chem. Phys.}\ }\textbf {\bibinfo {volume} {138}},\ \bibinfo {pages} {134115}
  (\bibinfo {year} {2013})}\BibitemShut {NoStop}%
\bibitem [{\citenamefont {Rubel}\ \emph {et~al.}(2021)\citenamefont {Rubel},
  \citenamefont {Tran}, \citenamefont {Rocquefelte},\ and\ \citenamefont
  {Blaha}}]{RUBELCPC21}%
  \BibitemOpen
  \bibfield  {author} {\bibinfo {author} {\bibfnamefont {O.}~\bibnamefont
  {Rubel}}, \bibinfo {author} {\bibfnamefont {F.}~\bibnamefont {Tran}},
  \bibinfo {author} {\bibfnamefont {X.}~\bibnamefont {Rocquefelte}},\ and\
  \bibinfo {author} {\bibfnamefont {P.}~\bibnamefont {Blaha}},\ }\href@noop {}
  {\bibfield  {journal} {\bibinfo  {journal} {Comput. Phys. Commun.}\ }\textbf
  {\bibinfo {volume} {261}},\ \bibinfo {pages} {107648} (\bibinfo {year}
  {2021})}\BibitemShut {NoStop}%
\bibitem [{\citenamefont {Schubert}\ \emph {et~al.}()\citenamefont {Schubert},
  \citenamefont {Kalantari}, \citenamefont {Lechner}, \citenamefont
  {Giesriegl}, \citenamefont {Nandan}, \citenamefont {Leiva}, \citenamefont
  {Kashiwaya}, \citenamefont {Sauer}, \citenamefont {Foelske}, \citenamefont
  {Ros\'{a}n}, \citenamefont {Blaha}, \citenamefont {Cherevan},\ and\
  \citenamefont {Eder}}]{SchubertAdM21}%
  \BibitemOpen
  \bibfield  {author} {\bibinfo {author} {\bibfnamefont {J.~S.}\ \bibnamefont
  {Schubert}}, \bibinfo {author} {\bibfnamefont {L.}~\bibnamefont {Kalantari}},
  \bibinfo {author} {\bibfnamefont {A.}~\bibnamefont {Lechner}}, \bibinfo
  {author} {\bibfnamefont {A.}~\bibnamefont {Giesriegl}}, \bibinfo {author}
  {\bibfnamefont {P.~S.}\ \bibnamefont {Nandan}}, \bibinfo {author}
  {\bibfnamefont {P.~A.}\ \bibnamefont {Leiva}}, \bibinfo {author}
  {\bibfnamefont {S.}~\bibnamefont {Kashiwaya}}, \bibinfo {author}
  {\bibfnamefont {M.}~\bibnamefont {Sauer}}, \bibinfo {author} {\bibfnamefont
  {A.}~\bibnamefont {Foelske}}, \bibinfo {author} {\bibfnamefont
  {J.}~\bibnamefont {Ros\'{a}n}}, \bibinfo {author} {\bibfnamefont
  {P.}~\bibnamefont {Blaha}}, \bibinfo {author} {\bibfnamefont
  {A.}~\bibnamefont {Cherevan}},\ and\ \bibinfo {author} {\bibfnamefont
  {D.}~\bibnamefont {Eder}},\ }\href@noop {} {}\bibinfo {note}
  {Unpublished}\BibitemShut {NoStop}%
\bibitem [{\citenamefont {Bader}(1985)}]{doi:10.1021/ar00109a003}%
  \BibitemOpen
  \bibfield  {author} {\bibinfo {author} {\bibfnamefont {R.~F.~W.}\
  \bibnamefont {Bader}},\ }\href@noop {} {\bibfield  {journal} {\bibinfo
  {journal} {Acc. Chem. Res.}\ }\textbf {\bibinfo {volume} {18}},\ \bibinfo
  {pages} {9} (\bibinfo {year} {1985})}\BibitemShut {NoStop}%
\end{thebibliography}%

\end{document}